\documentclass[%
 reprint,
superscriptaddress,
nofootinbib,
 amsmath,amssymb,
 aps,
]{revtex4-2}

\usepackage{graphicx}
\usepackage{dcolumn}
\usepackage{bm}
\usepackage{hyperref}

\usepackage{amsmath}
\usepackage{amsfonts}

	\usepackage{graphicx,graphics,xcolor,braket,amsmath,mathtools,subfigure,bm,bbm,tensor,array}
\usepackage{hyperref}
\usepackage{cleveref}
\usepackage{stmaryrd}
\usepackage{xfrac} 
\usepackage{multirow}
\usepackage{enumitem}
\usepackage{float}

\newcommand{\PDF}{{\rm PDF}}
\newcommand{\Normal}{{\mathcal N}}
\newcommand{\Prior}{{\mathcal P}}
\newcommand{\Expect}{{\rm E}}

\newcommand{\Tr}{{\rm Tr}}
\newcommand{\NLL}{{\mathcal L}}
\newcommand{\Pfail}{M}
\newcommand{\maj}{\rm maj}

\newcommand{\true}{\textsf{t}}
\DeclareMathOperator*{\argmin}{\arg\!\min}

\newcommand{\LS}[2]{{}^2{\textrm{#1}}_{#2}}

\begin{document}

\title{Optimised Bayesian system identification in quantum devices}

\author{\mbox{Thomas M.\ Stace}}
\author{\mbox{Jiayin Chen}}
\author{\mbox{Li Li}}
\author{\mbox{Viktor S.\ Perunicic}}
\author{\mbox{Andre R. R. Carvalho}}
\author{\mbox{Michael Hush}}
\affiliation{
 Q-CTRL, Sydney, NSW 2000, Australia}
\author{\mbox{Christophe H. Valahu}}
\author{\mbox{Ting Rei Tan}}
\affiliation{School of Physics, University of Sydney, NSW 2006, Australia}
\affiliation{ARC Centre of Excellence for Engineered Quantum Systems, University of Sydney, NSW 2006, Australia}
\author{\mbox{Michael J.\ Biercuk}}
\affiliation{
 Q-CTRL, Sydney, NSW 2000, Australia}
\affiliation{School of Physics, University of Sydney, NSW 2006, Australia}
\affiliation{ARC Centre of Excellence for Engineered Quantum Systems, University of Sydney, NSW 2006, Australia}

\begin{abstract}
Identifying and calibrating quantitative dynamical models for physical quantum systems is important for a variety of applications.  Here we present a closed-loop Bayesian learning algorithm for estimating multiple unknown parameters in a dynamical model, using optimised experimental ``probe'' controls and measurement.  The estimation algorithm is based on a Bayesian particle filter, and is designed to autonomously choose informationally-optimised probe experiments with which to compare to model predictions.  We demonstrate the performance of the algorithm in both simulated calibration tasks and in an experimental single-qubit ion-trap system. Experimentally, we find that with $60\times$ fewer samples, we exceed the precision of conventional calibration methods, delivering an approximately $93\times$ improvement in efficiency (as quantified by the reduction of measurements required to achieve a target residual uncertainty and multiplied by the increase in accuracy).  In simulated and experimental demonstrations, we see that successively longer pulses are selected as the posterior uncertainty iteratively decreases, leading to an exponential improvement in the accuracy of model parameters with the number of experimental queries.
\end{abstract}

\maketitle



Accurately and efficiently identifying parameters in quantitative dynamical models that describe quantum systems is an open challenge in the development of quantum technologies \cite{PhysRevA.91.022125}.  This is an important task broadly known as System Identification, which enables various applications including system tuneup, error-budgeting, and control design.  More specifically, in the context of quantum computing, possession of an accurate and predictive system model can be used to develop  high-fidelity error-robust quantum logic gates~\cite{PhysRevApplied.15.064054};  the accuracy of such models will be a key determinant~\cite{Mart_nez_Garc_a_2019} in our ability to reduce gate errors substantially below fault-tolerant error-correction thresholds~ \cite{10.5555/2011763.2011764,doi:10.1146_annurev-conmatphys-031119-050605}.

Increasing the accuracy of parameter estimates typically involves the performance of comprehensive - but resource intensive - experimental processes~\cite{Nielsen2021gatesettomography} which can conflict with the fundamental premise that in real experiments quantum measurements are generally expensive.  This is a consequence of both the exponential growth of the parameter-space with system size, and fundamental measurement-induced noise, requiring repeated state-preparation and evolution cycles to build up statistics of measurement outcomes.  Further, measurements in real quantum devices are typically up to an order of magnitude slower than  other quantum gates \cite{PhysRevLett.119.180501,Google2021,Pino:2021tf,Postler:2022uq,ETHsurface}.  Finding measurement-efficient routines enabling high-fidelity parameter estimates is therefore critical for the advancement of useful experimental methods in parameter estimation.  

Here we introduce a closed-loop learning-control algorithm which iteratively finds information-maximizing control pulses~\cite{NEURIPS2019_d55cbf21,pmlr-v108-foster20a} in order to improve the accuracy of uncertain model parameters, such as Hamiltonian coefficients, with high efficiency in the number of measurements employed. The approach we adopt adaptively adjusts an interrogating control pulse waveform (duration, amplitude, phase, etc) in a manner similar to techniques used for high-dimensional control-parameter optimisation~ \cite{PRXQuantum.2.040324}; here, informationally-optimised probe waveforms are chosen based on a user-defined system model and the prior state of knowledge of the system parameters in a Bayesian framework.  Importantly, this approach can accommodate realistic system constraints in the optimisation process, such as control band-limits, known linear or time-invariant transfer functions, and nonlinearities.  The algorithm tracks the distribution of model-parameter values with a sample population, which is filtered and repopulated using a particle filter as new experimental data becomes available.  This results in a sequence of posterior populations \cite{978374,PRXQuantum.3.020350} that incorporate the measured data and enable rapid convergence.  We demonstrate that this algorithm delivers an exponential reduction in parameter uncertainty over iterations in both simulated and experimental tests using trapped atomic ions.  In these tests the algorithm learns to increase the probe duration in order to optimise the trade-off of probe sensitivity and uncertainty due to aliasing, and fully utilises its freedom to design novel control probes for high-dimensional systems.

\begin{figure*}[t]
\begin{center}
\includegraphics[width=\textwidth]{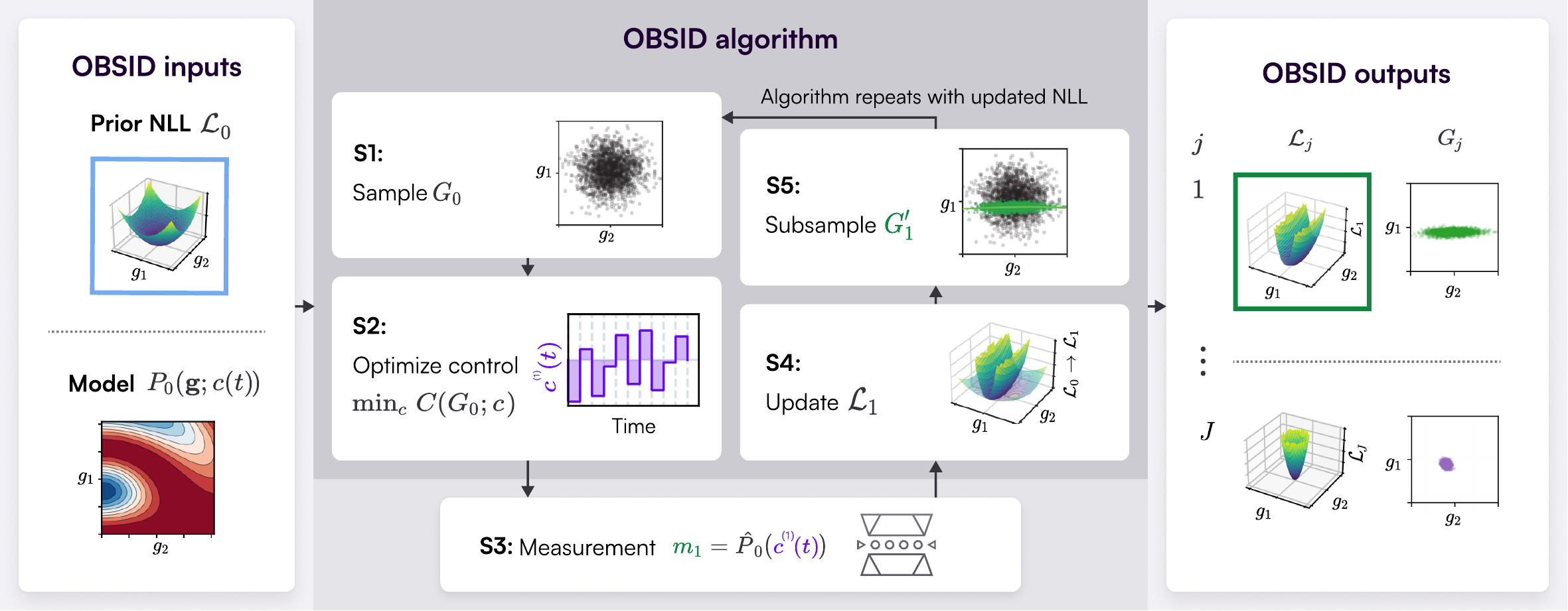}
\end{center}
\caption{
A loop of the OBSID algorithm, shown in the dark grey box.  The inputs to the OBSID algorithm are the  prior negative-log likelihood (NLL), ${\NLL}_{0}$ (blue square), representing the prior uncertainty over the system parameters \mbox{${\bf g}=\{g_1,g_2,...\}$}, and the system model.  The prior NLL is used in [S1] to sample a prior population of model parameters, $G_0$ (grey points).  [S2] An optimised control pulse $c^{(1)}(t)$ is found using the response $P_0({\bf g},c)$ predicted by the system model to maximise the anticipated parameter information gain, based on the cost, $C$, averaged over $G_0$.  [S3] The optimised control is used to drive an experiment.  The experimental measurement outcome, $m_1$, is used to  [S4] update the posterior NLL, \mbox{${\NLL}_{1}={\NLL}_{0}+\delta \NLL_1$}, and  [S5]  subsample a posterior population, \mbox{$G'_1\subset G_0$} (green points), which is used to compute the posterior sample mean $\bar {\bf g}_1$ and covariance matrix $\Sigma_1$. The updated NLL, ${\NLL}_{1}$ (green square), becomes the prior for the next iteration of OBSID (light arrow).  The output from the OBSID algorithm is the sequence of posterior NLLs, $\NLL_j$, and parameter populations, $G_j$, determined after each measurement.  The algorithm completes after it reaches a target accuracy, iteration limit $J_{\max}$,  or other termination criteria.   Further details are given in \Cref{sec:OBSID}.}
\label{fig:schematic}
\end{figure*}

In \Cref{sec:background} we introduce the context for system identification, including an illustrative two-parameter system model which also serves as a testbed for subsequent benchmarking.  We then use the Fisher Information \cite{kok_lovett_2010,PhysRevA.91.022125} to elucidate important factors in system identification of uncertain model parameters. \Cref{sec:OBSID} describes in detail our Optimised Bayesian System Identification (OBSID) protocol, illustrated in \Cref{fig:schematic}, which is the central innovation in this manuscript.  In \Cref{sec:benchmarking} we demonstrate the algorithm on simulated one- and two-qubit systems, described by models with up to five parameters.  We also validate the protocol in an experimental calibration of a ion trap system, and achieve parameter estimates in agreement with standard metrological approaches but requiring $\sim 20\times$ fewer experimental measurements.  We conclude with a \hyperref[sec:discussion]{discussion} of OBSID applications and usage.   

\section{System Identification Context}\label{sec:background}



Formally, by \textit{system identification}  we mean the measurement and calibration of initially uncertain parameters, ${\bf g}$, in a quantitative model for some experimental system.  This may include terms in a Hamiltonian, decay rates in a dissipative system, or calibration constants relating  control pulses produced by a signal generator to the field experienced by a physical system, including e.g.\ distortion or cross-talk.  


\begin{figure*}
\begin{center}
\includegraphics[]{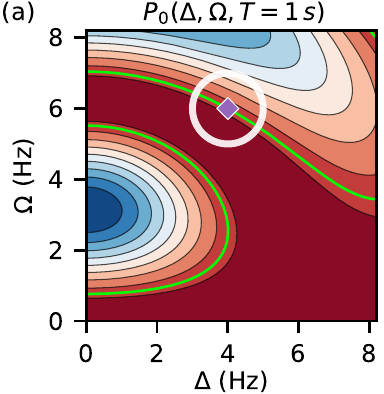}
\includegraphics[]{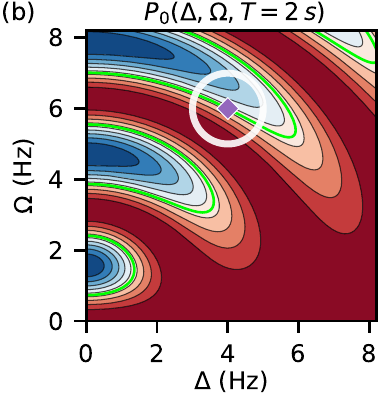}
\includegraphics[]{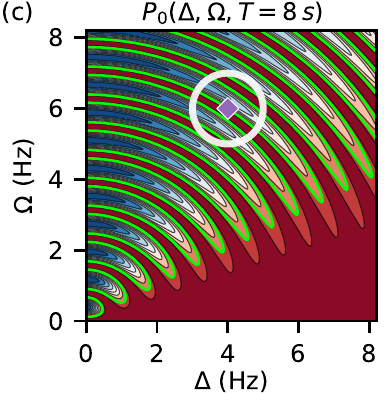}
{\raisebox{7mm}{
\includegraphics[]{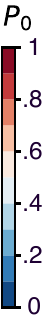}}}
\hspace{5mm}
\includegraphics[]{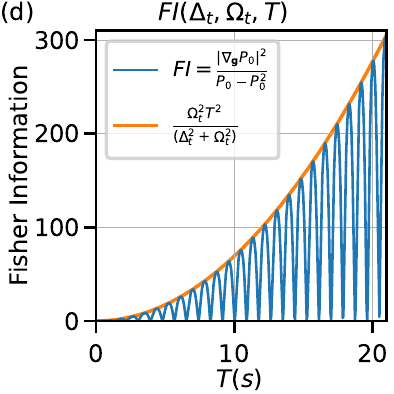}
\end{center}
\caption{Contours of the return probability landscape, $P_0(\Delta,\Omega;T)$, for the one-qubit, two-parameter model from \cref{eqn:H1q}, driven by a uniform (Rabi-type) control pulse $c(t)=1$ of duration (a) $T=1$,  (b) $T=2$, and (c) $T=8$ `seconds'; note that the units here are entirely nominal.  
In each plot, we indicate a specific point at ${\bf g}^{(1\rm q)}_\true=\{\Delta_\true,\Omega_\true\}$ Hz (purple diamond). 
For a system whose true parameters were ${\bf g}^{(1\rm q)}_\true$, the green contour cutting through ${\bf g}^{(1\rm q)}_\true$ represents the set of parameter values that would be consistent with an experimental measurement, $\hat P_0$, using a control pulse of the corresponding pulse duration, $T$.  The white \mbox{$1\sigma$-uncertainty} disk pictured is helpful for illustrating the competing effects that lead to an optimal pulse duration, as described in the main text.  
(d) Shows the Fisher information for this model, which grows as $T^2$ with an oscillatory modulation.
} \label{fig:paramscaling}
\end{figure*}


To make explicit the ingredients for system identification, we assume the following:
\begin{enumerate}[label=A{\arabic*}.,left=0cm]
    \item \emph{Initialisation} -- the experimental system can be reliably initialised into a specific reference  state $\ket{0}$, such as the ground state or a separable product state.
    \item \emph{Control fields} -- the system evolves under application of  externally specified time-dependent control fields, \mbox{${\bf c}=\{c_1(t),c_2(t),...\}$} of duration $T$, which can be modulated by a programmable signal generator.  For example, several qubits may be simultaneously addressed by several control fields.
    \item \emph{Measurements} --  projective measurements are performed in orthogonal subspaces $\{\Pi_0,\Pi_1,...\}$, returning empirical probabilities $\hat{\bf P}({\bf c})=\{\hat P_{{0}},\hat P_{{1}},...\}$ 
    and their corresponding uncertainties,  \mbox{${\bf \sigma}=\{ \sigma_{{0}}, \sigma_{{1}},...\}$}. We will assume that measurement outcomes are Normally distributed around their mean \mbox{$\hat P_{i}\sim \Normal_{P_{i},\sigma_{{i}}^2}$}.
     \item   \emph{Model} --  evolution of the system under investigation is well-described by a quantitative model with $p$ model parameters ${\bf g}=\{g_1,g_2,...,g_p\}$ that determine the evolution.  For a given choice of ${\bf g}$ and time-dependent control pulses ${\bf c}$, we use the model to compute probabilities ${\bf P}({\bf g},{\bf c})=\{ P_0, P_1,...\}$ for the subspaces $\{\Pi_0,\Pi_1,...\}$.  Comparing the experimental measurement $\hat{\bf P}({\bf c})$ with model predictions ${\bf P}({\bf g},{\bf c})$ provides  information about ${\bf g}$. 
    \item  \emph{Prior uncertainty} -- the model parameters have a prior uncertainty which we assume is a multi-variate Normal distribution,  ${\bf g}\sim\Prior_0= \Normal_{\bar {\bf g}_0,\Sigma_0}$, with prior mean, $\bar{\bf g}_0$ and covariance matrix, $\Sigma_0$. 
\end{enumerate}
Here, we write an abstract distribution over the $p$-dimensional  model parameter space as ${\bf g}\sim\Prior_j$, and the corresponding probability density  as $\PDF_j({\bf g})$. 

Assumptions A1 to A5 are practically reasonable, and are sufficient to define the goals of OBSID.  A key insight from these assumptions is that, since we have a choice about what probe-pulse to deliver to the experiment, there is an opportunity to optimise this choice of control in order to maximise the information gained about ${\bf g}$.  

Within OBSID (Fig~\ref{fig:schematic}), at each iteration of the algorithm, we numerically search for control pulses, ${\bf c}$ which maximise the anticipated information gain of proposed experiments, averaged over the prior parameter uncertainty.  We then apply the optimised control to the physical (or simulated) system and measure the response $\hat{\bf P}$.  This is compared with the model prediction $\bf P$, in order to update a posterior distribution over model parameters under a Bayesian framework.

We note that various of these assumptions can be relaxed if required.  For example mixed initial states in A1 and open evolution in A4 can be accommodated, albeit with additional computational cost.  In A3, generalised quantum measurements can be accommodated if necessary, as well as non-Normal distributions in A3 and A5.

\subsection{Illustrative model for probe-control optimisation}

We introduce an illustrative model that highlights the role of the control probe-pulse in the efficiency of the system-identification procedure. The physical model we build upon embodies many salient features in OBSID, and is also used later in some of the one- and two-qubit benchmarking results we report in \Cref{sec:benchmarking}. It is a two-parameter model, ${\bf g}^{(1\rm q)}=\{\Delta,\Omega\}$, for a single qubit driven by one control field, $c(t)$, with Hamiltonian
\begin{equation}
    H^{(1\rm q)}(t)=-\Delta\, Z/2 +\Omega\, \big(c(t) L+c(t) ^* L^\dagger\big)/2,
    \label{eqn:H1q}
\end{equation}
where $\Delta$  is  the detuning, $\Omega \,c(t)$ is the  Rabi frequency,  $\Omega$ is the  Rabi calibration factor, $Z$ is a Pauli operator, and $L$ is the corresponding lowering operator for a two-level system.  We note that $c(t)$ has arbitrary units specified at the output of a signal generator (e.g.\ `volts').  

For this single-qubit model, we assume that the qubit is initialised in the ground state of the undriven Hamiltonian, $\ket{0}$, and measured in the undriven energy eigenbasis $\Pi_0=\{\ket{0}\}$ and $\Pi_1=\{\ket{1}\}$, so that $P_0=|\langle 0|\psi(t)\rangle|^2$, and $P_1=1-P_0$.  Indeed, for the rest of this paper we will assume that system measurements are two-state projectors, in which we measure the \textit{return} probability, $P_0$, and the complement $P_1$. 


Using this model, Figures \ref{fig:paramscaling}(a-c) illustrate the prototypical system response, $P_0({\bf g}^{(1\rm q)}; c(T))$ which  reflects  assumptions A1 to A5.  Here $T$ is the control pulse duration, assuming a Rabi-type control pulse of unit amplitude, $c(t)=1$.  We plot the response for three different pulse durations, $T=1,2$ and 8 `seconds' (using a normalized energy basis), in each case starting from the ground state $\ket{0}$.  
Shown in \Cref{fig:paramscaling}(a-c)  as a purple diamond is a putative (but arbitrary) `true' system parameter, \mbox{${\bf g}^{(1\rm q)}_\true=\{\Delta_\true,\Omega_\true\}$ Hz}.  For a given pulse, measurement of the system response will yield a measured probability $\hat P_0$, indicated by the green contours that pass through ${\bf g}^{(1\rm q)}_\true$, up to measurement noise.  Illustrating assumption A5, the white $1\sigma$-uncertainty disk in each panel represents prior uncertainty in the parameters.  

This  model gives us a simple example to illustrate the effect of different pulse choices (e.g.\ durations) on the posterior uncertainty.  For the shortest pulse shown in \Cref{fig:paramscaling}(a), $P_0$ is a relatively slowly varying function over parameter space.  For longer pulses (\Cref{fig:paramscaling}(b,c)), $P_0$ becomes increasingly oscillatory. The slowly varying response landscape of the shortest pulse has a unique contour cutting the uncertainty disk, highlighted in green.  However, because the response is slowly varying (i.e.\ shallow gradient), the posterior distribution will be loosely clustered around the green cutting contour, with relatively large spread off the contour.  That is, the response has low sensitivity, so the posterior distribution will have relatively large posterior uncertainty. 

By contrast, for the long-duration control pulse shown in \cref{fig:paramscaling}(c) the rapidly-varying landscape has high sensitivity to the parameter values, and consequently the posterior distribution would be relatively tightly clustered around the consistent cutting contours (green lines).  However, since there are multiple consistent contours cutting through the uncertainty disk, the posterior distribution will still have wide support across the disk. For an experimentalist this is a manifestation of the familiar $\mod{2\pi}$ uncertainty associated with a periodic function such as the measured population under Rabi oscillations.

The response at the intermediate duration shown in \cref{fig:paramscaling}(b)  with $T=2$, is ideal (in this example) . Its sensitivity is as large as possible, giving a  posterior distribution that is tightly clustered around the green contour, while still having a unique contour cutting the disk. This illustrates a pulse for which the anticipated posterior uncertainty will be small. Qualitatively, the intermediate pulse duration represents a `Goldilocks' choice with highest information gain about the parameters, given the indicated uncertainty region.  

More generally, the tradeoff identified above between the response sensitivity and the density of consistent contours within a prior uncertainty region illustrates the possibility for varying the duration (and shape) of the control signal, ${\bf c}$, to maximise the information gain about the model parameters ${\bf g}$.  Iterating, as the uncertainty region shrinks, the control pulse should be adapted accordingly.

\subsection{Sensitivity, uncertainty and Fisher information}\label{sec:FI}
To help formalise these observations, we introduce the  Fisher information (FI), which is a widely used proxy for the sensitivity of a probability distribution with respect to its parameters \cite{kok_lovett_2010}. Typically, larger FI corresponds to higher sensitivity with respect to the model parameters; and correspondingly  higher potential information gain about unknown parameters.
 The FI  also provides intuition about the `geometry' of information in the $p$-dimensional model parameter space.  

For a multi-parameter problem, with the two-state response probabilities $\{P_0,P_1\equiv1-P_0\}$, the Fisher information Matrix (FIM) \cite{PhysRevA.91.022125} is given by 
\begin{equation}
    {\rm FIM}({\bf g};{{\bf c}})=\vec f({\bf g}; {{\bf c}}) \boxtimes \vec f({\bf g}; {{\bf c}}),\label{eqn:FIM}
\end{equation}
where $\vec f({\bf g};{{\bf c}})= (P_0-P_0^2)^{-1/2}\nabla_{\bf g} P_0({\bf g};{{\bf c}})$ is the FI eigenvector, and $a\boxtimes b$ denotes the outer product of  $a$ and $b$.
For a two-state  response, the FIM is a rank-1 matrix. The FI  is the unique non-zero eigenvalue of the FIM
\begin{equation}
    {\rm FI}({\bf g}; {{\bf c}})=|\vec f({\bf g}; {{\bf c}}))|^2.\label{eqn:FI}
\end{equation}
For later discussion, we refer to the unit-vector 
\begin{equation}
    \tilde f({\bf g};{\bf c})=\vec f({\bf g};{\bf c})\big/|\vec f({\bf g};{\bf c}) |= \nabla_{\bf g} P_0\big/|\nabla_{\bf g} P_0|\label{eqn:ID}
\end{equation}
 as the \emph{information direction}. This defines the direction at each point in parameter space  along which measurement of $\hat P_0({{\bf c}})$ reveals information, and is locally orthogonal to contours of constant $P_0$.  
It follows that for a $p$-dimensional model space, we need at least $p$ independent control fields, $\{{\bf c}^{(1)},{\bf c}^{(2)},...,{\bf c}^{(p)}\}$ to generate a spanning set of information directions, $\{\tilde f({\bf g};{\bf c}^{(1)}),...,\tilde f({\bf g};{\bf c}^{(p)})\}$, that constrain the posterior distribution over all $p$ parameters.  These statements can be generalised to multi-state measurement outcomes.

The FI is a function of the control pulse envelope, and it is generally very strongly dependent on the overall control pulse duration, $T$.  This is illustrated in \Cref{fig:paramscaling}(d), which shows the FI as a function of $T$, for the single-qubit model generated by \cref{eqn:H1q} with $c(t)=1$.  The envelope  grows as $T^2$, with an oscillatory modulation.  

The quadratic growth of FI with pulse duration is a fairly generic property \cite{PhysRevA.92.032124}. For unitary evolution, when $g_i\in{\bf g}$ are energetic scales in a Hamiltonian, then 
\begin{equation}
    {\rm FI}({\bf g}, T) = T^2 \vartheta({\bf g},T),\label{eqn:FIscaling}
\end{equation}
where $\vartheta$ is a bounded and generally oscillatory function of pulse duration and the Hamiltonian parameters (see the supporting discussion in  \Cref{AppendixFI}).  It follows that large FI will be obtained by increasing $T\rightarrow\infty$.  However, given the foregoing `Goldilocks' argument that the optimal \emph{finite} duration, $T^{({\rm opt})}<\infty$, should depend on the prior parameter uncertainty, we see that maximising FI alone is not sufficient to define informationally-optimal control pulses given finite prior uncertainty.  


This qualitative discussion gives rise to a rough estimate of the optimal control pulse duration, assuming that parameters in $\bf g$ are energy scales in a Hamiltonian.  Firstly, we assume that the prior parameter uncertainty is specified by a covariance matrix, $\Sigma$, which determines an uncertainty ellipse in parameter space \cite{10.2307_2348711} (e.g.\ illustrated in \Cref{fig:paramscaling} as a white disk). Then the length of the major axis of the ellipse is given by the largest eigenvalue,
$\lambda^{\rm maj}$, of the deviation matrix $\Sigma^{1/2}$, which corresponds to the maximally uncertain combination of model parameters.  Secondly, both $P_0$ and the FI oscillate in parameter space with a period $|\delta {\bf g}|\sim1/ T$ (and taking $\hbar=1$ so that energy and time are reciprocal units).  

Optimal pulses tend to be those for which the oscillation period in parameter space matches the major uncertainty, i.e.\ $|\delta {\bf g}|\approx\lambda^{\rm maj}$, corresponding to the `Goldilocks' situation depicted in  \Cref{fig:paramscaling}(b).  It follows that the duration of the optimal pulse is $T^{\rm(opt)}\sim1/\lambda^{\rm maj}\gtrsim \Tr({\Sigma^{-1/2}})$. This is an important estimate for two reasons: it gives intuition for how optimal pulse durations relate to the size of the parameter uncertainty ellipse, and it gives a practically useful order-of-magnitude for initialising numerical optimisation over $T$ in OBSID.



\section{Optimised Bayesian system identification protocol}\label{sec:OBSID}

\subsection{OBSID algorithm overview}
The OBSID protocol starts from a prior uncertainty in the model parameters, $\Prior_0$, and with the aid of the system model that computes $\bf P$, finds a control pulse that is likely to optimally improve our knowledge of the model parameters.  We encode the anticipated knowledge gain from an arbitrary pulse in terms of a cost function, $C$.  Using this optimised control, an experiment is performed, and the measurement results are incorporated into a posterior distribution, $\Prior_1$, over model parameters.  Iterating this yields a sequence of probability distributions $\Prior_j$, quantified by their density functions, $\PDF_j({\bf g})$, that iteratively and autonomously localises the model parameters.

A loop of the OBSID algorithm is illustrated in \Cref{fig:schematic}.    The inputs to the algorithm are the system model with which to compute $P_0({\bf g}, {\bf c})$, and the initial prior distribution, $\Prior_0$ over parameters ${\bf g}$, represented as a negative-log likelihood (NLL) function, \mbox{$\NLL_0({\bf g})=-\ln(\PDF_0({\bf g}))$}, shown as 2-dimensional surface inside the blue square in \Cref{fig:schematic}.   Numerically, we track the sequence of posterior distributions, $\Prior_j$, using a Bayesian population filter \cite{978374}, which form the output of the algorithm.

The algorithm itself, shown inside the dark-grey box in \Cref{fig:schematic}, loops over the following sample--optimise--measure--inference steps, initialised with $j=1$, and terminating after reaching a desired accuracy, iteration limit $J$,  or other terminal conditions:
{
\begin{enumerate}[label=S{\arabic*}.,left=0cm]
    \item  {Sample} a prior population, $G_{j-1}=\{{\bf g}_s\}_{s=1}^S$, from the prior distribution \mbox{$\PDF_{j-1}({\bf g})=\exp(-\NLL_{j-1}({{\bf g}}))$}, so that ${\bf g}_s\sim \mathcal{P}_{j-1}$.
    \item {Minimise} a population cost function $C(G_{j-1},{\bf c})$ over pulses ${\bf c}$, to find an optimal pulse \mbox{${\bf c}^{(j)}=\argmin_{\bf c}C(G_{j-1},{\bf c})$}.  The optimisation uses the model predictions, $P_0({\bf g}_s, {\bf c})$, at each ${\bf g}_s\in G_{j-1}$, and also model gradients $\nabla_{\bf g} P_0({\bf g}_s, {\bf c})$.
    \item {Run} an experiment with the optimal pulse, ${\bf c}^{(j)}$, to measure $m_j=\hat P_0( {\bf c}^{(j)})$, with  uncertainty $\sigma_j$ determined by the measurement statistics.
    \item {Update} the posterior NLL over \mbox{${\bf g}_s\in G_{j-1}$}, so that $\NLL_{j}({{\bf g}_s}|m_j)=\NLL_{j-1}({{\bf g}_s})+\delta \NLL_{j}({{\bf g}_s},m_j)$, where 
    \begin{align}
    \delta \NLL_j({{\bf g}},m_j)&=-\ln\big(\delta L_j (m_j|{{\bf g}}\big),\nonumber\\
    &\equiv\big(P_0({\bf g}, {\bf c}^{(j)})-m_j\big)^2/(2 \sigma_j)^2,\label{eqn:deltaL}
    \end{align}
   assuming that \mbox{$\delta L_j(m|{{\bf g}}\big)=\Normal_{ P_0({\bf g},{\bf c}^{(j)}),\sigma_j}(m)$}. 
    \item {Return}:
    \begin{itemize}[leftmargin=0.6cm,labelsep=0.3cm]
         \item  a posterior likelihood subsample $G'_{j}\subset G_{j-1}$, i.e.\ retain $g_s\in G_{j-1}$ with probability ${\delta L}_{j}({{\bf g}_s},m_j)$,
         \item  the posterior sample mean, \mbox{$ \bar {\bf g}_{j}(m_j)=\Expect_S({\bf g}|m_j)$},  
         \item the posterior sample covariance,
   \mbox{$
     \Sigma_{j}(m_j)=\Expect_S\big(({\bf g}-\bar {\bf g}_j)\boxtimes({\bf g}-\bar {\bf g}_j)\big|m_j \big)$},
    \end{itemize}
where the sample expectation of $x$ conditioned on $m$ is  $\Expect_S(x|m)=\sum_s x_s\, \widetilde{\delta L}_{j}({{\bf g}_s},m)$, and we have defined the sample-normalised likelihood  
    \begin{equation}
        \widetilde{\delta L}_{j}({{\bf g}},m)={\delta L}_{j}(m|{{\bf g}})/N_j(m),\label{eqn:likelihood}
    \end{equation}
    where \mbox{$N_j(m)=\sum_s{\delta L}_{j}(m|{{\bf g}_s})$} and $\delta L_j$ is defined implictly in \cref{eqn:deltaL}.  
    \item {Increment} $j$ or terminate.
\end{enumerate}}

The iterative outputs of the loop constitute a sequence of posterior quantities  $\{\bar {\bf g}_{j}(m_j),\Sigma_{j}(m_j), G_j\}_{j=1}^J$, which increasingly localise the system parameters as measurement data is accumulated and processed.

\subsection{Comments on implementation}

Step S1 is implemented using importance sampling to sample a population from  the prior NLL, $\NLL_{j-1}$ \cite{tokdar2010importance}.  This is depicted as a population point cloud in \cref{fig:schematic}.    Importance sampling is implemented by generating a trial presample $T_{j-1}=\{{\bf g}_s^T\}$ from a proposal distribution, which we take to be the multi-Normal distribution so that \mbox{${\bf g}^T_s\sim \mathcal{T}_{J-1} = \Normal_{\bar {\bf g}_{j-1}, \Sigma_{j-1}}$}, and then assigning an importance score   
\begin{equation}
    i_s\propto \PDF_{j-1}({\bf g}_s^T)/\Normal_{\bar {\bf g}_{j-1}, \Sigma_{j-1}}({\bf g}_s^T) \in(0,1].
\end{equation}  
We retain ${\bf g}_s^T\in G_{j-1}\subseteq T_{j-1}$ with probability $i_s$.  If necessary, we repeat this procedure until the resampled posterior population reaches a predefined  size,  typically a few thousand sample points.

Step S2 finds a control pulse that optimises the anticipated information gain, as quantified in a cost function, $C$.  In \Cref{sec:cost} below, we describe cost functions that are suitable for OBSID. 
We have implemented the OBSID algorithm, including the cost function evaluation, in a graph-based machine-learning (ML) environment \cite{Ball_2021,boulder_opal2}, building on \textit{TensorFlow} \cite{tensorflow2015-whitepaper}, which facilitates automatic differentiation for the purpose of efficient computation and gradient-based optimisation of $C$. This environment is designed for automated quantum control and learning applications, and it efficiently scales to handle controls with many optimizable parameters.  It also straightforwardly accommodates soft constraints in the cost function that encode desirable properties of control pulses, such as limited bandwidth, slew-rate, etc. This computational approach delivers greater flexibility to efficiently search the control space relative to previous discrete optimisations constrained to changing only the number of repetitions of a fixed probe control in related experiments~\citet{PRXQuantum.3.020350}.

Step S3 requires an interface to a physical (or simulated) experiment.  The interface should communicate pulse specifications provided by OBSID to the experimental apparatus' control system, which then implements the pulse with high-fidelity on the physical target system.  Care needs to be taken that the control system does not produce unmodelled distortion (e.g.\ nonlinearities, cross-talk or dead time in buffers) in the pulse as received by the physical target.  If such effects are present, they should be included in the system model's parameterization.  The interface waits for the experiment and measurement process to complete, and then receives the measurement outcome $m_j$ along with the standard measurement error, $\sigma_j$. 

Step S4 is a direct applications of Bayes' rule to compute the posterior NLL given the measurement outcome, $m_j$. Internally, the NLL is stored as an interpolating function which can be evaluated anywhere inside the convex-hull of the sample points in $G_{j-1}$; this implements a consistent probability density estimator for amortising the computation of previous predictive calculations and experimental measurements, as described in \cite{NIPS2013_7f53f8c6}.  The RHS of \cref{eqn:deltaL} implements assumption A3 that the measurement statistics are Normally distributed.  However, this can be generalised to other statistical distributions, e.g.\ the binomial distribution for a finite number of Bernoulli trials, if necessary.

Step S5 uses rejection sampling  \cite{tokdar2010importance} based on the acceptance likelihood, $\delta L_j$, to generate a posterior sub-population $G_j'$, shown as green points in \cref{fig:schematic}, and computes the posterior sample mean and covariance matrix, which are reported as part of the output of the protocol.

Lastly, the posterior NLL from S4 becomes the prior NLL to start the next iteration of the OBSID loop, illustrated in the green square in \cref{fig:schematic}.

\subsection{Cost functions}\label{sec:cost}

The cost function that is minimised in S2 is a critical part of the protocol.  We have implemented two suitable cost functions.  The first, $C_{\rm APC}$, is based on the \textit{Anticipated Posterior Covariance} (APC), which was described in  \cite{PRXQuantum.3.020350}.
The second, $C_{\rm MFI}$, is based on a modified Fisher Information that penalises highly oscillatory responses, and is described in \Cref{AppendixFI}.

In practice we find that both cost functions work comparably well.  However, $C_{\rm APC}$ is more straightforward to describe and is more generalizable, so for benchmarking OBSID in later sections, we only present results based on $C_{\rm APC}$, which we describe presently.  The existence of at least two cost functions is conceptually useful as a foil to abstract the OBSID protocol from any specific choice of cost function. This abstraction opens avenues to finding other cost functions with desirable properties, such as reduced computational overhead \cite{NEURIPS2019_d55cbf21,pmlr-v108-foster20a}, or enhanced robustness. We comment on alternative cost functions in \Cref{AppendixAltCost}.


The motivation for the APC is that for a given pulse ${\bf c}^{(j)}$, the experiment will return some measurement outcome, $m_j=\hat P_0( {\bf c}^{(j)})$ with statistical uncertainty $\sigma_j$.  Of all the prior parameter samples, ${\bf g}_s \in G_{j-1}$, the ones that are consistent with $m_j$ are those for which 
\begin{equation}
    P_0({\bf g}_s, {\bf c}^{(j)})\approx m_j\pm \sigma_j.\label{eqn:approxP}
\end{equation} 
More precisely, under assumption A3, the relative likelihood of a prior sample ${\bf g}_s\in G_{j-1}$ point to be represented in a posterior population, $G_{j}$ is given by 
\begin{equation}
    {\delta L_j}({{\bf g}_s},m_j)=e^{-(P_0({\bf g}_s, {\bf c}^{(j)})-m_j)^2/(2 \sigma_j)^2},
\end{equation}
consistent with \cref{eqn:approxP}. Rejection sampling with this relative likelihood produces a statistically consistent posterior subsample, $G_{j}'$.  The posterior sample covariance, $\Sigma_j$, depends implicitly  on ${\bf c}^{(j)}$, and so the APC cost function is designed to reward control pulses that minimise $\Sigma_j$, in some measure.

Since we must choose ${\bf c}^{(j)}$ prior to the measurement which determines $\Sigma_j$, we average the posterior sample covariance over all possible measurement outcomes \cite{NEURIPS2019_d55cbf21} to produce the \emph{anticipated} posterior covariance
\begin{equation}
    \Xi_j
    =\int dm \,Q_j(m){\Sigma}_{j}(m),\label{eqn:Xi}
\end{equation} 
where $Q_j(m)=N_j(m)/\int dm' \, N_j(m')$ is the likelihood of anticipated measurement outcomes, and   \mbox{$N_j(m)$} is defined in Step S5.   
The APC matrix, $\Xi$, which was introduced in \cite{PRXQuantum.3.020350}, is positive definite and depends implicitly on the pulse ${\bf c}^{(j)}$.


A well-chosen pulse is one for which the eigenvalues of $\Xi$ are small, with a particular bias towards minimising the largest eigenvalue, which is the major posterior uncertainty.   A convenient cost function that captures this is 
\begin{align}
    C_{\rm APC}({\bf c}^{(j)})&=\Tr\big(A\cdot\Xi_j
    \big),\nonumber\\
    &=\int dm \,Q_j(m)
    \,\Tr\big(A\cdot {\Sigma}_{j}(m)\big).\label{eqn:ave} 
\end{align}
where $A$ is a diagonal preconditioning matrix that rescales parameter uncertainties if required (e.g.\ if the model parameters carry different units, or are of greatly differing orders of magnitude).  For benchmarking  in the rest of this paper, we take $A=\mathbb{I}$.  In practice, we approximate the nested integrals in \Cref{eqn:Xi,eqn:ave} over $m$ and $m'$ as a discretised sum over the range of possible measurement outcomes represented across the prior parameter sample population $G_{j-1}$.


\section{Performance benchmarking}\label{sec:benchmarking}

In this section, we benchmark the performance of OBSID in estimating Hamiltonian parameters for both simulated and experimental systems.  The scope for parameterizing control pulses is essentially unlimited, for example,  superpositions of smooth basis functions, or piece-wise continuous (PWC) functions.  In these demonstrations we use PWC controls with a user-defined segment count, and with variable segment amplitudes and durations, so that the overall pulse shape and duration are included in the optimisation. 
%
In this section, we first demonstrate OBSID on the simulated single-qubit calibration of a two-parameter Hamiltonian model described by \cref{eqn:H1q}.  We then validate this with the experimental calibration of an ion-trap system.  Finally, we  demonstrate the calibration of a simulated two-qubit system characterised by a five-parameter model, with two independent control fields.

\subsection{Simulated single-qubit parameter estimation}

We begin by demonstrating parameter estimation for the pair ${\bf g}^{(1\rm q)}$ used to specify the model introduced in \cref{eqn:H1q} via numerical simulation of the physical model. The simulator calculates the time evolution of the system under an arbitrary control and generates measurement statistics, $m_j$ and $\sigma_j$, replicating experimentally-realistic finite sampling errors.  We assume that the true parameter values are \mbox{${\bf g}^{(1\rm q)}_\true=\{\Delta_\true,\Omega_\true\}=\{4,6\}$ Hz}, (noting that in all simulations units are entirely nominal).  We assume independent Normally-distributed priors  \mbox{$ \Delta_0\sim4.1\pm0.5 \text{ Hz}$} and \mbox{$ \Omega_0\sim6.2\pm0.5 \text{ Hz}$} (here \mbox{$x\sim \mu\pm \sigma$} signifies that \mbox{$x\sim\Normal_{\mu,\sigma^2}$}).   

\subsubsection{Automated, optimised Rabi-Ramsey measurements}\label{sec:rabi}

\begin{figure}
\begin{flushleft}
{\includegraphics
{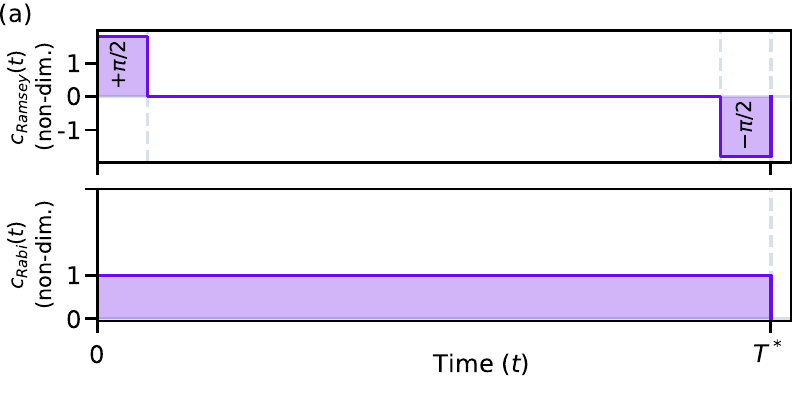}\hfill}

\vspace{1mm}
{
\includegraphics[]
{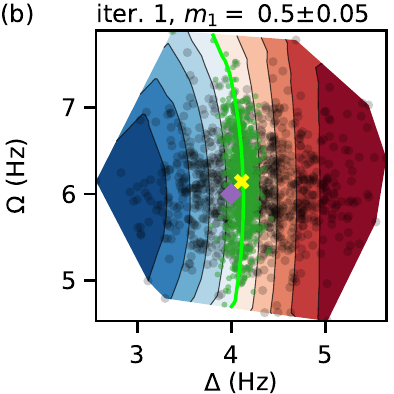}
\includegraphics[]
{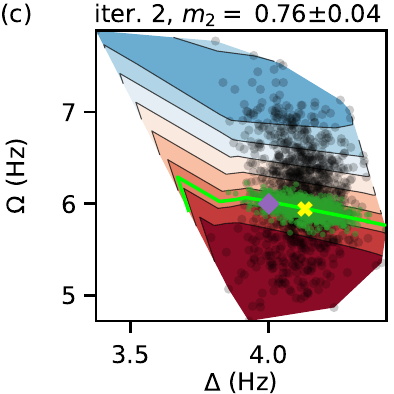}\hfill
{\raisebox{8mm}{
\includegraphics[]{figs/labelled_legend.pdf}}}
}
{\includegraphics[]
{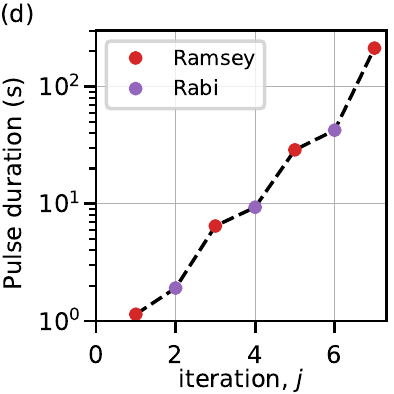}
\includegraphics[]
{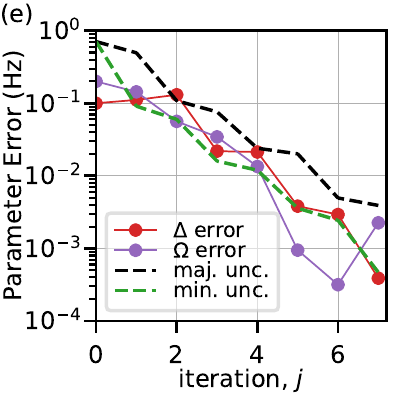}
\hfill}
\end{flushleft}
\caption{
Results of simulated single qubit calibration using conventional `Ramsey' and `Rabi' type control pulses. (a) shows the pulse envelopes for the two control types, in which the  duration $T^{(j)}$ is optimisable.  The first and last segments in the Ramsey control pulse are (approximate) $\pm\pi/2$ pulses. 
Panels (b) and (c) show the first two iterations of the prior population (black circles) and posterior population (green dots), superimposed on the predicted response, $P_0$ (contours), for the optimally chosen pulse.  The green contour is the measured value $m_j=\hat P_0$ (from simulation).  The posterior mean (yellow cross) and true parameter value (purple diamond) are also shown.  The  posterior population (green) at iteration 1 becomes the prior population for iteration 2 (black). (d) Pulse durations and types selected to optimally estimate single qubit parameters, and (e) the iterative improvement in the absolute parameter errors, $|\bar{\bf g}_j-{\bf g}_\true|$, and the uncertainty characterised by the eigenvalues of $\Sigma_{j}^{1/2}$.
} 
\label{fig:RabiRamseyResults}
\end{figure}

Measuring ${\bf g}^{(1\rm q)}=\{\Delta,\Omega\}$ is a standard single-qubit calibration scenario, which requires a minimum of two independent parameterisations.  This is conventionally accomplished by a combination of experiments that use a uniform Rabi-type pulse, 
\begin{equation}
   c_{\rm Rabi}(t)=\Theta(T_{\rm Rabi}-t),
\end{equation}
where $\Theta(x)$ is the unit-step function,  and experiments that use a Ramsey-type pulse,
\begin{equation}
    c_{\rm Ramsey}(t)=\tfrac{\pi}{2\Omega}\big(\delta(t)-\delta(t-T_{\rm Ram})\big),
\end{equation} 
which begins and ends with approximate $\pm\pi/2$ pulses.
These two control parameterisations are shown schematically in \Cref{fig:RabiRamseyResults}(a). Rabi-type experiments  are sensitive to the combination \mbox{$(\Omega^2+\Delta^2)^{{1}/{2}}\equiv \Omega_{\rm eff}$}, 
while Ramsey-type experiments are sensitive to the free evolution determined by $\Delta$.  In combination,  the interleaved experiments are sufficient to localise both parameters.

We first illustrate the performance of OBSID in a simulated scenario that optimises this conventional single-qubit tuneup procedure.  At each iteration, OBSID  autonomously optimises over Rabi or Ramsey pulse-types, choosing the pulse-type and duration that minimises $C_{\rm APC}$.  The simulated experiment is then addressed with the optimal pulse, and returns \mbox{$m_j=\hat P_0(c^{(j)})$} and the simulated sample uncertainty $\sigma_j$.  

\Cref{fig:RabiRamseyResults}(b) and (c) show the outputs from the first two iterations of the OBSID loop, with the prior population at each iteration shown as grey points, and the posterior population after importance resampling as green points.  Dark contours show the predicted response, $P_0$, over the parameter space, and the contour consistent with the `measured' value $m_j$ returned from the simulated experiment is shown in green.  Because this is a simulation, we know the true parameter values ${\bf g}^{(1\rm q)}_\true$ (shown as a purple diamond), which are contained inside the convex-hull of the posterior population (green points), illustrating the ability of OBSID to rapidly converge on an effective estimate of ${\bf g}^{(1\rm q)}$.  

\Cref{fig:RabiRamseyResults}(d) shows the optimal duration and pulse type chosen at each OBSID iteration.  The algorithm autonomously determines a procedure which alternates between Ramsey and Rabi-type pulses. The effect of the alternating  pulse-types chosen by OBSID is seen in \cref{fig:RabiRamseyResults}(b) and (c).  The first iteration selected a Ramsey-type pulse, which principally localises $\Delta$ and whose insensitivity to $\Omega$ is evident in the orientation of the contours of \cref{fig:RabiRamseyResults}(b); the second iteration chose a Rabi-type pulse that localises along a contour of constant $\Omega_{\rm eff}$, which is locally a linear combination of $\Delta$ and $\Omega$ in the neighbourhood of the `true' value, ${\bf g}^{(1\rm q)}_\true$.  

The estimated posterior uncertainty after each iteration is measured by the largest (major) eigenvalue of the posterior sample deviation matrix, \mbox{$\lambda_j^{\rm maj}={\rm MaxEval}[\Sigma_j(m_j)^{1/2}]$}.  This is shown in \cref{fig:RabiRamseyResults}(e) (dashed black line), and is a statistical upper bound for the uncertainty of all marginals.  We also plot the absolute error between the estimated population mean and the known values used in the simulation, shown as points.  We expect that the absolute error in the parameters, $|\bar{\bf g}_j-{\bf g}^{(1\rm q)}_\true|\lesssim \lambda_j^{\rm maj}$. This is borne out in practice.

Returning to \Cref{fig:RabiRamseyResults}(d) we also observe that the optimised pulse duration grows by an approximately constant factor from iteration to iteration, consistent with the Fisher information tending to grow with $T$.  This results in an overall exponential growth in the pulse duration with iteration count, $j$. Thus the optimal pulse duration varies approximately inversely with the major uncertainty, as discussed earlier.  This is a general characteristic of all simulations we have performed.

\subsubsection{Automated, optimised arbitrary pulse measurements}

\begin{figure}
\begin{flushleft}
{\includegraphics
{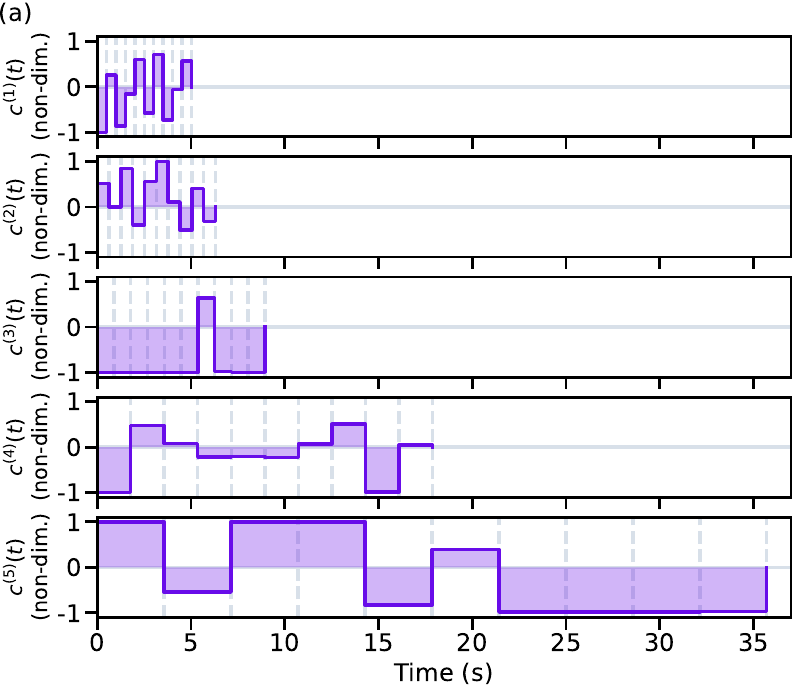}\hfill}

\vspace{1mm}
{

{\includegraphics[
]{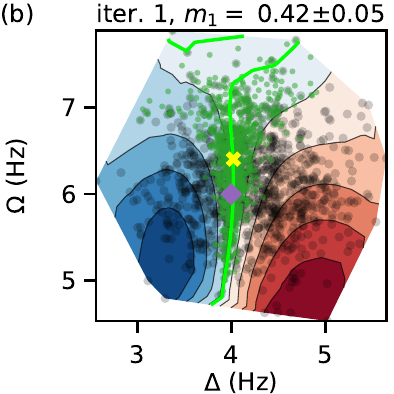}
{\includegraphics[
]{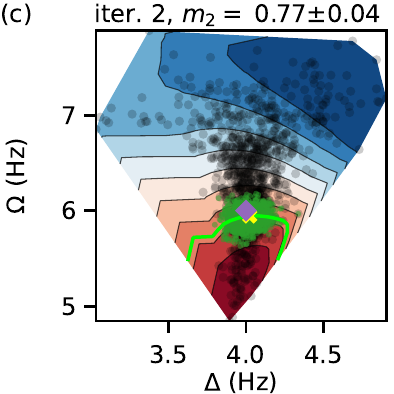}\hfill
{
\raisebox{8mm}{\includegraphics[]
{figs/labelled_legend.pdf}}}
}
}}
{\includegraphics[]
{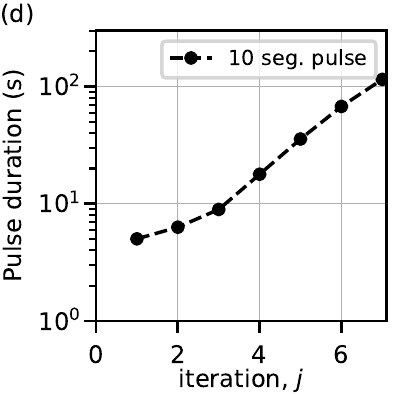}
\includegraphics[]
{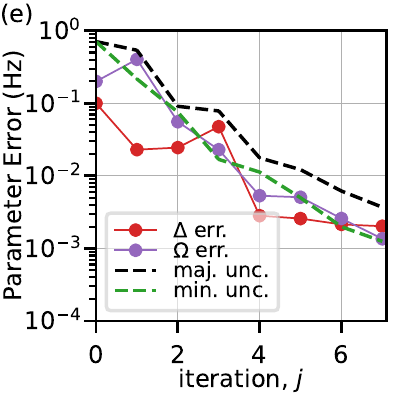}
\hfill}
\caption{
Results of simulated single qubit calibration using fully optimised 10-segment piecewise-continuous pulses. (a) The sequence of pulses generated in each iteration of OBSID, in which the segment amplitudes and the total pulse duration $T^{(j)}$ are all optimised. 
(b) and (c) show populations for the first two iterations, showing prior  (grey points) and posterior  (green points) populations, superimposed on the predicted response, $P_0$ (interpolated contours), for the optimally chosen pulses. 
(d) Pulse durations for each iteration, and (e)  the iterative improvement in the parameter error and uncertainty.
} 
\label{fig:Arb2Results}
\end{flushleft}
\end{figure}

The example optimisation over Rabi and Ramsey pulses uses predefined control envelopes which have well-understood sensitivity to the model parameters.  However, the general approach in OBSID is capable of finding good control pulses directly, without relying on intuition or predefined control types.  This becomes important for parameter estimation in multi-dimensional systems which are complex enough that well-understood calibration and control pulse families are not available.  

We demonstrate this capability leveraging the same simulation, but now using a PWC control pulse whose segment amplitudes and total pulse duration are independently optimised.  To demonstrate the flexibility of OBSID, we show the results using 10-segment pulses (using fewer segments also works in practice).   During optimisation, the segment amplitudes are constrained to the real interval $[-1,1]$, and the maximum duration of each pulse is constrained by \mbox{$T^{(j)}\leq 2T^{(j-1)}$}, so that the overall duration does not grow too quickly. \Cref{fig:Arb2Results}(a) shows the pulses chosen by the OBSID protocol for the first five iterations.  Clearly the pulse durations tend to grow with iteration.

\Cref{fig:Arb2Results}(b) and (c) show the prior and posterior sample populations, along with the model-predicted response in the first two OBSID iterations.  \Cref{fig:Arb2Results}(d) shows the typical exponential growth in overall pulse duration, and (e) shows the major uncertainty, and the absolute parameter errors.  These outcomes are qualitatively the same as in \cref{fig:RabiRamseyResults} for the Rabi-Ramsey calibration discussed above.

Again, we observe that the OBSID algorithm selects exponentially growing pulse durations, $T^{(j)}$, with iteration count (\cref{fig:Arb2Results}(d-e)).  Further, the contours of the predicted response, $P_0$, which depends on $c^{(j)}$ characteristically rotate in parameter space from one iteration to the next: contours and the posterior population in the first iteration are broadly aligned to the $\Omega$-axis, while for the second iteration they are more aligned with the $\Delta$-axis.  This behaviour, in which sequential iterations produce roughly orthogonal  `cuts' through parameter space, is a consequence of the APC cost function that minimises the major uncertainty.  

The pulses selected by the OBSID algorithm achieve performance that is marginally better than results from using Rabi-Ramsey control pulses.  Specifically, comparing panels (d) and (e) in  \cref{fig:RabiRamseyResults} and \cref{fig:Arb2Results}  shows that after several iterations, the arbitrary pulse optimisation procedure returns pulses of somewhat shorter duration while maintaining similar accuracy in the resulting parameter estimates.  Unlike the Rabi-Ramsey sequences, the effect of the 10-segment PWC pulses are not readily interpretable except by direct simulation. 

\subsection{Experimental single-qubit parameter estimation in a trapped-ion system}\label{sec:expQCLdemo}

Having benchmarked the OBSID protocol for a single qubit using simulated experiments, we now validate it using a real trapped-ion experimental system. The qubit is encoded in the $\LS{S}{1/2}$ electronic ground state of a $^{171}{\rm Yb}^+$ ion, where we assign the qubit states $\ket{F=0, m_F=0} \equiv \ket{0}$ and $\ket{F=1, m_F=0}\equiv\ket{1}$. Considering a 12.64 GHz microwave field near resonance with the $\ket{0}\leftrightarrow\ket{1}$ transition, the system's dynamics are described by the same two-parameter Hamiltonian of \cref{eqn:H1q}. The detuning $\Delta$ corresponds to the frequency difference between the microwave control signal and the qubit's resonance frequency, while $\Omega$ is the calibration factor relating the actual Rabi frequency experienced by the ion to the microwave driving voltage amplitude from a signal generator, so it has units of `Hz/Volt'. 

The experimental apparatus consists of a linear Paul trap held at room temperature \cite{Milne2020}. 
The degeneracy of the $F=1$ Zeeman levels is lifted by applying a static magnetic field of 0.44 mT. The native, uncompensated qubit coherence time is measured to be \mbox{$T_2^* = 8.7(7)$ s} with a Ramsey-type experiment. A set of laser beams addressing the $\LS{S}{1/2}\rightarrow\LS{P}{1/2}$ transition near 369.5 nm are used for Doppler cooling, state preparation and measurement \cite{Olmschenk2007}. State detection is performed by discriminating between the differing number of scattered photons collected on an avalanche photodiode for each of the two qubit states, when turning on the 369.5 nm laser. 

The microwave control field is obtained from a synthesis chain in which a direct digital synthesizer (DDS, AD9910) is mixed with an arbitrary waveform generator (AWG, Keysight M8190A). The resulting signal is up-converted to 12.64 GHz. The DDS has a frequency resolution of 0.23 Hz and the clock speed of the AWG is 5.8 GHz; they both have am amplitude resolution of 14 bit. The 12.64 GHz signal is amplified and delivered to the ion by an antenna positioned 2.37 cm from the trap centre. The amplifier chain in the synthesizer system has been independently calibrated to assure linearity between the amplitude specified in a pulse sequence and the amplitude experienced by the ion. 


We use two-stage calibration measurements, implemented with DDS, to determine the best initial estimate of the `true' physical parameters which will later serve as `ground-truth' for the OBSID results. The true detuning $\Delta_\true$ is determined using 150 samples of a standard Ramsey-locking method \cite{Peik2005} with up to a 50 ms Ramsey interrogation time. The Rabi frequency $\Omega_\true$ is measured by fitting resonant Rabi oscillations over approximately 28 Rabi cycles.  The oscillations were sampled with 300 uniformly-spaced samples in time, so that there were about eleven samples per Rabi cycle.  Each data point in these experiments used between 50 and 100 individual, binary-valued single-qubit  fluorescence readouts to estimate $\hat P_0$.  
In total, 450 experimental measurement samples were taken (i.e.\ 150 for Ramsey, and 300 for Rabi), giving calibrated values \mbox{$\Omega_\true=1249.1\pm0.1$ Hz/Volt}, and \mbox{$\Delta_\true=500\pm0.4$ Hz}, where the uncertainty is given as the standard error.

\begin{figure}[t!]
\begin{center}
{\includegraphics[
]{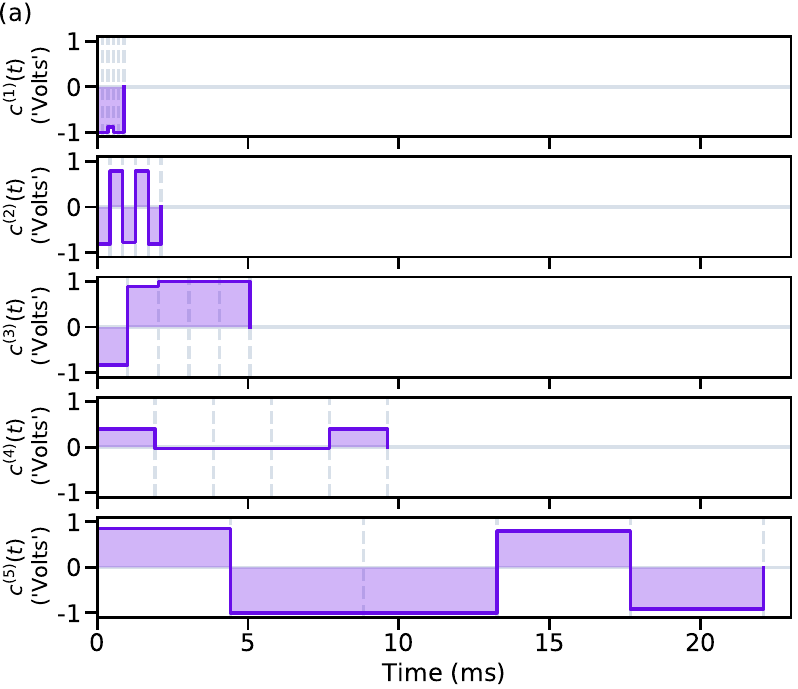}\hfill}
\vspace{2mm}

{\includegraphics[
]{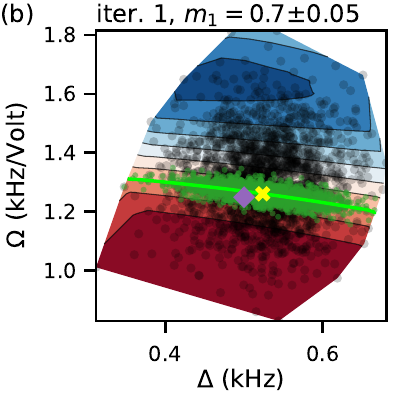}
\includegraphics[
]{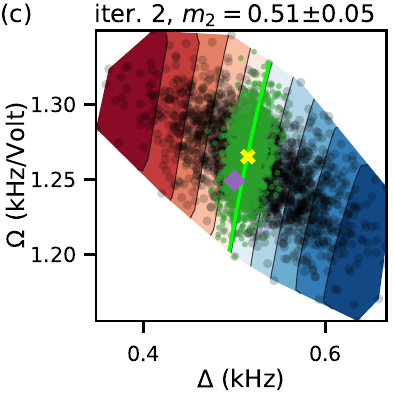}\hfill
{
\raisebox{8mm}{\includegraphics[]
{figs/labelled_legend.pdf}}}
}
\vspace{-3mm}
{\includegraphics[
]
{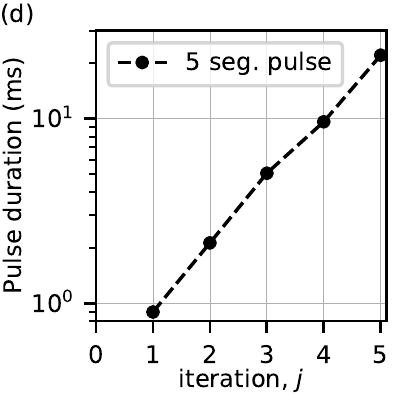}
\includegraphics[
]
{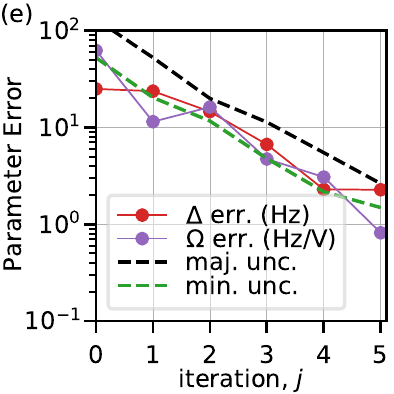}
\hfill}
\caption{
Experimental OBSID results in a single-ion qubit calibration, using optimised 5-segment pulses. (a) The sequence of pulses generated in the first five iterations of OBSID, in which the segment amplitudes and the total pulse duration $T^{(j)}$ are all optimised. 
(b) and (c) show populations for the first two iterations, showing prior  (grey points) and posterior  (green points) populations, superimposed on the predicted response, $P_0$ (interpolated contours), for the optimally chosen pulses. 
(d) Pulse durations for each iteration, and (e) the  iterative improvement in the parameter accuracy.  The floor in the absolute errors  is set by the inferred uncertainty in the conventional calibration methods, which has a standard error of  $\pm0.3$ Hz, for both $\Delta$ and $\Omega$.
} 
\label{fig:QCLResults}
\end{center}
\end{figure}

To initialise the OBSID protocol, we assume uncorrelated, Normally-distributed priors for each parameter, with \mbox{$\bar \Delta_0\sim525\pm52.5\text{ Hz/Volt}$} and \mbox{$\bar \Omega_0\sim1311\pm131.1\text{ Hz}$}.  This was chosen to reflect a scenario in which the prior standard-deviation is $10\%$ of the prior mean parameter values.  The prior uncertainty is large enough that it includes the pre-calibrated, `ground-truth' parameter values within a region of high probability density, and so is statistically consistent with the conventionally-calibrated system parameters.  Such a circumstance is conventionally encountered experimentally in which a `coarse' spectroscopy scan is followed by experimental fine tuning to estimate experimental parameters. The OBSID waveform is generated with an AWG.

\Cref{fig:QCLResults}(a) shows the first five PWC control pulses chosen by OBSID. The control pulse profiles specify the amplitudes input to  the AWG  to modulate the experimental drive, at a fixed carrier frequency.

Figures \ref{fig:QCLResults}(b) and (c) show the prior and posterior population of parameters for the first two OBSID iterations. As before, the posterior population mean (yellow cross) converges to the `ground truth' (purple diamond).   \Cref{fig:QCLResults}(d) shows the characteristic growth in control pulse duration, by roughly an order of magnitude over five iterations, consistent with the simulated results in \cref{fig:Arb2Results}(d).  

Lastly, \Cref{fig:QCLResults}(e) illustrates the convergence between the OBSID posterior and the ground truth values, along with the estimated major uncertainty from the population covariance matrix.  Again, we see consistency between the internal estimate of the major parameter uncertainty and the error relative to ground truth.  Given the accuracy of the conventional calibration measurements, the floor in each of the absolute errors is the same as the calibrated uncertainty of $0.1$ Hz and $0.4$ Hz for both $\Omega$ and $\Delta$.  

We see from  \Cref{fig:QCLResults}(e) that the major uncertainty estimate from OBSID reaches $\pm2.6$ Hz after $j=5$ samples. This is a statistical upper bound on the absolute error, which is $2.2$ Hz and $0.8$ Hz for both $\Omega$ and $\Delta$. The total iteration number, $j$, is limited by the AWG memory in our system due to a high sample rate, and does not pose a fundamental limit for the protocol. We compare OBSID to the conventional method of simultaneously estimating $\Omega$ and $\Delta$, fitting a detuned Rabi flop, in order make a quantitative comparison of protocol efficiency. We sample 300 equally spaced points up to the OBSID's longest duration, $22.1$ ms, and using the same number of experimental shots at each sample as in OBSID. The resulting uncertainties are $3.4$ Hz and $7.9$ Hz for $\Omega$ and $\Delta$, which is greater than the absolute error obtained by OBSID for a $60\times$ larger number of samples ($300$ vs $5$).  Overall this shows that for a comparable number of samples, and using the smaller residual uncertainty on $\Omega$ as a bound, OBSID is approximately $93\times$ more efficient than standard parameter estimation techniques


The experimental results shown in \cref{fig:QCLResults} are qualitatively indistinguishable from the simulated results shown in \cref{fig:Arb2Results}.  In both simulation and experiment, the OBSID protocol treats the simulation or experimental system as a black-box probability generator, and chooses informationally-optimised pulses based on the model and the sample population. This experimental demonstration of OBSID illustrates one of its main  strengths: OBSID does not require prior knowledge of `good' pulses to use in the calibration procedure.  This will become increasingly important in larger systems, as we treat next. 


Beyond the demonstrations in \cref{fig:QCLResults} we have experimentally implemented other parameterisations of control pulses for use with OBSID in this experimental system.  We present summary results for experimental implementations of these alternative pulse parameterisations in \Cref{app:QCL_other_pulses}:
\begin{itemize}[left=0.0cm,labelsep=0.2cm]
    \item The Rabi-Ramsey pulse parameterisation described in \cref{sec:rabi}, and illustrated schematically in \Cref{fig:RabiRamseyResults}(a).  OBSID autonomously selects the optimal pulse type and pulse duration at each iteration.  
    
    \item Optimised `bang-bang' control, which uses PWC pulses alternating between `on' (unit magnitude) and `off' (zero magnitude) segments.  OBSID independently optimises each of the segment durations, as well as the carrier phases, $e^{i\phi(t)}$, during the `on'-segments.   This pulse parameterisation has the advantage that, since the `on'-pulse carrier magnitude is constant, linearity of the transfer chain from the AWG to the target ion is not required.  
\end{itemize} 
In both cases, the results appear qualitatively the same as in Figures \ref{fig:RabiRamseyResults}, \ref{fig:Arb2Results}, and \ref{fig:QCLResults}.

\begin{figure*}[ht!]
\hfill
    \begin{minipage}[lt]{8.3cm}
    \begin{flushleft}
        \includegraphics[ trim=0 17 0 0,clip 
        ]{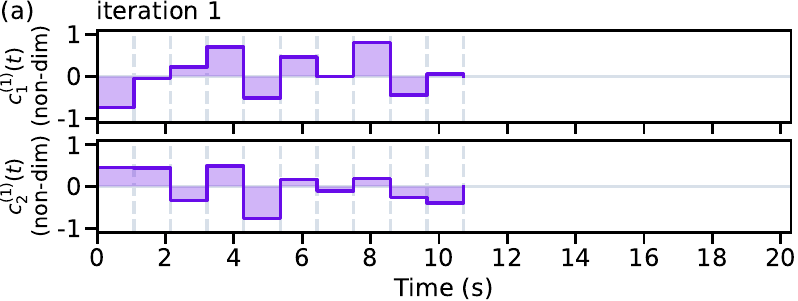}
        
        \vspace{1.7mm}
        \includegraphics
         [ trim=0 17 0 0,clip]{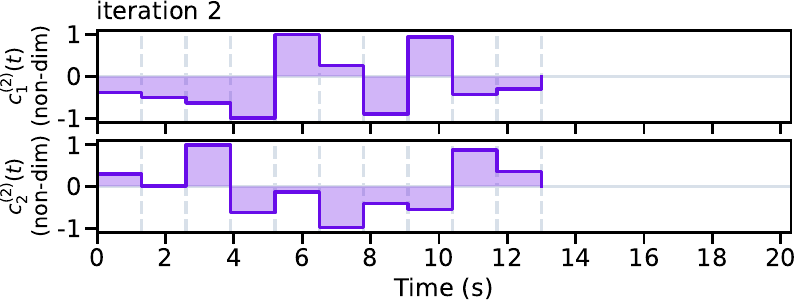}
         
         \vspace{1.7mm}
         \includegraphics[ trim=0 0 0 0,clip]{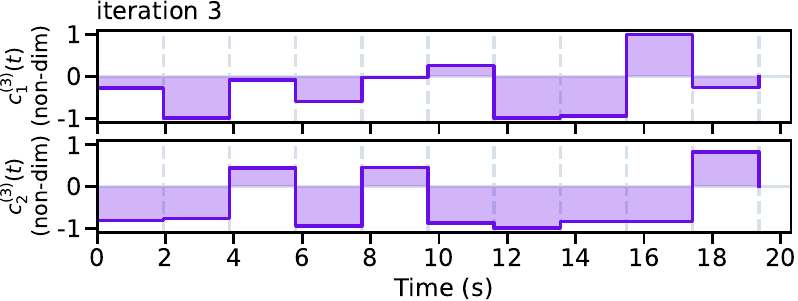}
         \end{flushleft}
\end{minipage}
\hfill
\begin{minipage}[lt]{8.9cm}
\begin{flushleft}
{\includegraphics[
]{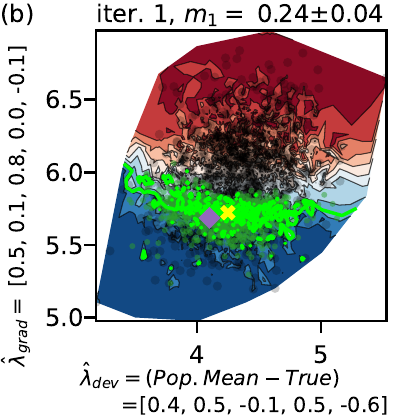}\hspace{2mm}
\includegraphics[
]{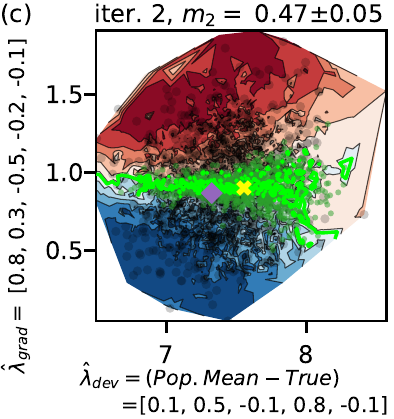}\hfill
{
\raisebox{10mm}{\includegraphics[]
{figs/labelled_legend.pdf}}}
}

\vspace{-2.5mm}
{\includegraphics[]
{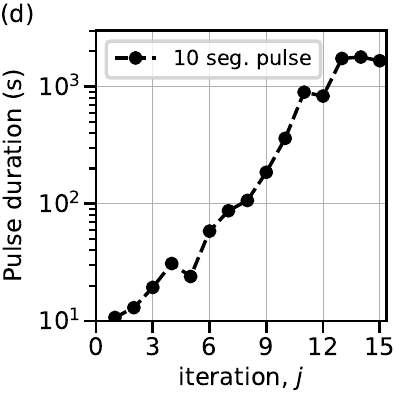}\hspace{2mm}
\includegraphics[]
{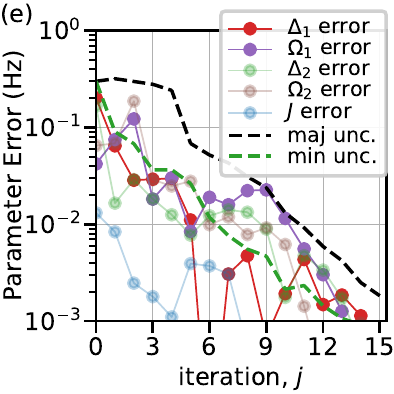}
\hfill}
\end{flushleft}
\end{minipage}
\hfill
\caption{
Results of SID on a simulated 2-qubit system with 5 parameters and 2 control fields.  (a) The sequence of optimised control pulse amplitudes, $c^{(j)}_1(t)$ and $c^{(j)}_2(t)$, applied to each qubit, for the first 3 iterations of OBSID.  (d) and (e) show scatter plots and contours of the response, $P_0$, as a function of parameters.  To represent the points in 5-dimensional parameter space,  we project onto two composite coordinates, defined by the unit vectors $\hat \lambda_{\rm dev}$  along the deviation between the true value and the population mean [horizontal axis], and $\hat \lambda_{\rm grad}$  along the direction of steepest gradient of $P_0$ [vertical axis]. Both unit vectors vary with iteration count, $j$, indicated by the numerical arrays shown in the axes labels. (f) shows that pulse durations tend to grow with iteration, and (g) shows that  uncertainties, estimated from the 5-dimensional population covariances and absolute errors both improve consistently over iterations.  We have highlighted the error in $\Delta_1$ and $\Omega_1$ relative to the simulated `true' value; fainter curves show other parameter errors evolve similarly.
}
\label{fig:2QubitResults}
\end{figure*}

\subsection{Simulated two-qubit parameter estimation}

We next demonstrate the operation of the OBSID protocol for a more complicated simulated five-parameter system, consisting of two coupled qubits, each of which is driven by an independent control field, specified by  time-dependent control fields $ c_q(t)$, for $q\in\{1,2\}$.  The Hamiltonian for this system is chosen to be
\begin{equation}
    H^{(2\rm q)}={\sum}_{q}
    H^{(1\rm q)}_q(t)+J (L_1^\dagger L_2+L_2^\dagger L_1),
\end{equation}
where $H^{(1\rm q)}$ was introduced in \cref{eqn:H1q}. The Hamiltonian is characterised by the 5 parameters
\begin{equation}
    {\bf g}^{(2\rm q)}= \{\Delta_1,\Omega_1,\Delta_2,\Omega_2,J \},\label{eqn:gt2q}
\end{equation}
with detunings, $\Delta_q$, Rabi frequencies, $\Omega_q \,c_q(t)$, and the inter-qubit coupling strength $J$.    We note in passing that this Hamiltonian could be derived from a system of two qubits driven by control fields that have a common carrier frequency \cite{PhysRevLett.107.080502,PhysRevA.97.012119}, however we use it here simply as a multi-parameter, multi-qubit demonstration of the SID technique.  

As in the single-qubit case, we assume that the qubits are initialised in the ground state of the undriven Hamiltonian, $\ket{\psi(0)}=\ket{0,0}$.  We assume a 2-state measurement process that  discriminates between the initial state, $\Pi_0=\{\ket{0,0}\}$, and the complementary subspace $\Pi_1=\{\ket{0,1},\ket{1,0},\ket{1,1}\}$, to yield the return probability, $P_0=|\langle 0,0|\psi(t)\rangle|^2$. This measurement model represents incomplete information for the two-qubit system, since only the return probability is distinguished.  This can be generalised to include an informationally-complete measurement, which will be the subject of future work.

For the purpose of simulation, we take the true parameter values to be ${\bf g}^{(2\rm q)}_\true=\{4.1,5.5,4,6,0.5\}\text{ Hz}$, and assume a prior deviation matrix given by \mbox{$\Sigma_0^{1/2}={\rm diag}\{0.3,0.3,0.3,0.3,0.03\}\text{ Hz}$}.  We initialise the OBSID prior with a mean value sampled from \mbox{$\bar{\bf g}_0\sim \Normal_{{\bf g}^{(2\rm q)}_\true,\Sigma_0}$}, and then set the prior to be \mbox{$\Prior_0=\Normal_{\bar{\bf g}_0,\Sigma_0}$}, that is, the relative uncertainty in each parameter is around 7\% of the true value.  

\Cref{fig:2QubitResults} shows the results obtained using OBSID for the 2-qubit model, using PWC control pulses with 10 segments.   \Cref{fig:2QubitResults}(a) shows the optimised pulse pairs, $c^{(j)}_1(t)$ and $c^{(j)}_2(t)$, chosen by OBSID for the first three iterations. In the pulse parameterisation shown here, at each iteration, the duration of the two control pulses are constrained to be the same.  This sequence of control-pulse pairs shows that OBSID is capable of optimising both control pulses simultaneously.  This would be particularly important in cases where calibrating cross-talk is necessary \cite{Milne2020}. 

Figures \ref{fig:2QubitResults}(b) and (c) show the prior and posterior population samples obtained during the first two iterations of OBSID. To partially visualise the five-dimensional parameter space, we project onto two particular unit-vectors. For the vertical axis of the projected plots, the  direction is chosen to align with the maximum average information direction,  $\hat\lambda_{\rm grad}\propto E_S(\tilde f({\bf g};{\bf c}^{(j)}))$, where the information direction, $\tilde f$ is defined in \cref{eqn:ID}. The unit vector defining the horizontal axis is chosen to align with the maximum deviation between the population mean and the true value used in the simulation, $\hat\lambda_{\rm dev}\propto \bar{\bf g}_j-{\bf g}^{(2\rm q)}_\true$.  These unit-vectors evolve over OBSID iterations, and we list the unit-vector coefficients in the axes labels with respect to the parameter ordering given in \cref{eqn:gt2q}.  

\Cref{fig:2QubitResults}(b) and (c) illustrate that, as for the two-parameter case, OBSID tends to choose control pulses whose response, $P_0$,  varies by a large amount along some direction in parameter space, and without oscillations in that direction.   Here, due to the dimensionality of the parameter space the axes in these graphs correspond to nontrivial cuts rather than the parameters themselves in previous figures (see caption). Importantly, the true value is located within the convex hull of the prior and posterior populations.  Further, the posterior distributions projected in \Cref{fig:2QubitResults}(b) and (c) are distinctly non-Gaussian.

\Cref{fig:2QubitResults}(d) again shows the tendency of the optimal control pulses to become longer with iteration count, which leads to a corresponding reduction in the estimated uncertainty in the parameters, shown in \Cref{fig:2QubitResults}(e).  The largest [smallest] estimated uncertainty is computed from the major [minor] eigenvalue of the population deviation matrix.  The major uncertainty is seen to  provide a good approximate bound on the absolute parameter error, 
$|\hat\lambda_{\rm dev} |$, shown as points in panel (e).  For clarity, in \cref{fig:2QubitResults}(e), we highlight the error in parameters $\Delta_1$ and $\Omega_1$, with other parameter errors indicated by fainter lines.  Experimentally, the true parameter values and absolute errors are unknown, but we reiterate that the posterior sample deviation provides a reliable estimate for the parameter uncertainty.

\section{Discussion}\label{sec:discussion}

These results, including simulated and experimental tests, illustrate that OBSID is able to efficiently converge on estimates that closely approximate the true parameter values for a system.  Comparing the single-qubit, two-parameter results in \cref{fig:Arb2Results}(e) with the two-qubit, five-parameter results in \cref{fig:2QubitResults}(e), we see empirically that the number of iterations to improve the parameter uncertainty by a fixed factor has scaled with the number of unknown parameters, $p$.  

A heuristic argument supporting this observation is as follows.  Suppose that the prior uncertainty density was a spherically-symmetric distribution in the $p$-dimensional model parameter space.  Each iteration of OBSID compresses the prior distribution in some direction in parameter space by a factor $\alpha<1$.  (For example, by inspecting  panels (b) and (c) in all the OBSID outputs given earlier, the compression of the posterior population samples relative to the prior samples shows that $\alpha\approx0.1$ to 0.2.)  Since OBSID preferentially compresses the most uncertain direction, after $p$ iterations, it will have compressed the spherical prior approximately symmetrically in all $p$ dimensions, and each will be compressed by approximately the same factor, $\alpha$.  So after $p$ iterations, the initially spherical uncertainty prior will have been compressed approximately uniformly.  Iterating, it follows that the number of iterations to uniformly shrink the prior uncertainty scales with $p$.


OBSID as we have designed and implemented it is extremely flexible.  For standard calibration tasks, it is capable of optimising the conventional Rabi-Ramsey based calibration schedule for a single qubit which alternates between localising the detuning and the Rabi frequencies. As shown in \cref{fig:RabiRamseyResults}(b) and (c), these choices correspond to cuts through parameter space which are approximately oriented orthogonal to the corresponding parameter at each iteration.  This has merit for `human-interpretability', in situations where it is desirable to calibrate model parameters individually.  In this case, the pulse shapes in \cref{fig:RabiRamseyResults}(a) that isolate a given parameter are known from conventional practice.  For more general problems, the APC cost-function could be straightforwardly adapted to select pulses that preferentially constrain parameters along the named parameter axes, that is, to include a `parameter-selective' preference in the cost function.  However, for highest efficiency, OBSID accesses the full $p$-dimensional parameter space, as illustrated in \cref{fig:Arb2Results} and \cref{fig:2QubitResults}.  In doing so, OBSID tends to introduce non-zero covariances, which correlates the uncertainty of different parameters.

The power of OBSID comes from increasing the control pulse duration, $T$, as the estimated uncertainty decreases.  For suitable control-pulse parameterisations, this allows the difference in the dynamical states generated by different prior model parameter samples to grow in the system Hilbert space, amplifying the effect of small parameter uncertainty into large differences in Hilbert space evolution without introducing additional uncertainty due to oscillatory behavior.  

We understand this through the Fisher information (FI), which  ideally  grows as $T^2$.  In such cases, for $r$ repetitions of a measurement, the parameter uncertainty scales as $\sigma\propto 1/\sqrt{{\rm FI}\, r} \sim 1/(T\sqrt{r})$.  Since the total accumulated measurement time scales as $\tau_{\rm meas}\sim T r$, the overall measurement time resource is best utilised by choosing as long a pulse duration $T$ as possible.  OBSID does this dynamically: it increases $T$ as quickly as possible, implicitly constrained by the prior uncertainty at each iteration.

The computational cost of optimising population-averaged model predictions over control pulses is non-trivial.  Our current implementation of OBSID builds on \textit{TensorFlow} \cite{tensorflow2015-whitepaper}, a graph-based toolbox for machine learning optimisation.  This gives it great flexibility in searching for optimal pulses, and automatically distributes highly parallelizable tasks across computational resources \cite{Ball_2021,boulder_opal2}.  For example, we have run the optimisation tools on systems ranging from few-core laptops to high-performance workstations with many cores, with run-times that scale with the resource availability.  However, the optimisation task is necessarily computationally demanding, typically taking tens of seconds to run on a 64-core workstation running at 4 GHz.  Improving this performance by using alternative less-intensive cost functions is a subject of recent \cite{NEURIPS2019_d55cbf21,NIPS2013_7f53f8c6} and ongoing research.


Using OBSID effectively requires some care.  It relies on statistical sampling, and so has some possibility to fail in any give run of system identification.  This was noted in \cite{PRXQuantum.3.020350}, and can be mitigated by including more sample points, at the cost of slower optimisation or greater parallel resources.  OBSID depends on a numerical simulation of a model, and so is subject to a computational cost that scales with the computational complexity of evaluating the model.  This makes it unsuitable for large-dimensional Hilbert spaces, unless there are efficient approximation schemes available for simulating the dynamics of specific systems.

There is a more subtle issue, related to \emph{statistical} and \emph{structural identifiability} \cite{sontag2013mathematical,10.1093/bioinformatics/btp358}, which determine whether OBSID succeeds.  Not all control parameterizations are suitable for effectively extracting model parameters with arbitrarily high accuracy.  For example, if the control pulse parameterization has fewer free control parameters than there are model parameters to fit, we expect OBSID will not be able to identify all system parameters, as described in \Cref{sec:FI}.  In this situation, the span of the information direction unit vectors does not cover the parameter space, and the control parameterization will not be able to structurally identify the system model parameters.   In practice, when a system is not structurally identifiable for a specific control parameterization, the major uncertainty reported by OBSID will stall after several iterations, ceasing to improve with additional iterations.  In \Cref{sec:identifiability} we describe an example of a control parameterization that is not structurally identifiable and which illustrates this stalling behaviour.   In practical usage of OBSID, simulations can be used to establish whether a system is structurally identifiable under a chosen control parameterization, before applying it to an experimental system.

\section*{Conclusions}
We have described and demonstrated a general purpose optimised Bayesian system identification protocol which uses model-based optimisation over a statistically sampled prior population to choose control pulses that optimally extract information about unknown system parameters.   Once the optimised pulse is applied experimentally, the system uses standard Bayesian update and resampling methods to produce a posterior distribution that is consistent with the new measurement information.  

The implementation we have demonstrated in this paper is autonomous and hardware agnostic, and requires-only that the system is well described by a quantitative numerical or analytical model.  The typical performance of the scheme is demonstrated in several simulated systems, as well as in a single-ion experiment.  We have shown experimentally that OBSID uses far fewer experiments to reach the same level of accuracy as conventional calibration methods; in the example presented, OBSID used just 5 different experiments to reach the accuracy of conventional single-ion calibration methods, which used 300 different experiments.  

There are a variety of areas to explore, including quantifying the robustness and reliability of using OBSID in experimental systems with many parameters, and in open quantum systems with dissipative effects.  There are also opportunities for optimising the algorithmic and computational run-time performance using alternative cost functions. 

\begin{acknowledgments}
This work was supported by the U.S. Army Research Office/Laboratory for Physical Sciences Grant \mbox{No.\ W911NF-21-1-0003}. Experimental results reported here were undertaken at the University of Sydney and supported by the Intelligence Advanced Research Projects Activity Grant \mbox{No.\ W911NF-16-1-0070}, the Australian Research Council Centre of Excellence for Engineered Quantum Systems Grant No.\ CE170100009, and a private grant from H.\ and A.\ Harley.  T.\ R.\ Tan was supported by the Sydney Quantum Academy Postdoctoral Research Fellowship. 
\end{acknowledgments}

\bibliography{paper.bib}

\begin{thebibliography}{33}%
\makeatletter
\providecommand \@ifxundefined [1]{%
 \@ifx{#1\undefined}
}%
\providecommand \@ifnum [1]{%
 \ifnum #1\expandafter \@firstoftwo
 \else \expandafter \@secondoftwo
 \fi
}%
\providecommand \@ifx [1]{%
 \ifx #1\expandafter \@firstoftwo
 \else \expandafter \@secondoftwo
 \fi
}%
\providecommand \natexlab [1]{#1}%
\providecommand \enquote  [1]{``#1''}%
\providecommand \bibnamefont  [1]{#1}%
\providecommand \bibfnamefont [1]{#1}%
\providecommand \citenamefont [1]{#1}%
\providecommand \href@noop [0]{\@secondoftwo}%
\providecommand \href [0]{\begingroup \@sanitize@url \@href}%
\providecommand \@href[1]{\@@startlink{#1}\@@href}%
\providecommand \@@href[1]{\endgroup#1\@@endlink}%
\providecommand \@sanitize@url [0]{\catcode `\\12\catcode `\$12\catcode
  `\&12\catcode `\#12\catcode `\^12\catcode `\_12\catcode `\%12\relax}%
\providecommand \@@startlink[1]{}%
\providecommand \@@endlink[0]{}%
\providecommand \url  [0]{\begingroup\@sanitize@url \@url }%
\providecommand \@url [1]{\endgroup\@href {#1}{\urlprefix }}%
\providecommand \urlprefix  [0]{URL }%
\providecommand \Eprint [0]{\href }%
\providecommand \doibase [0]{http://dx.doi.org/}%
\providecommand \selectlanguage [0]{\@gobble}%
\providecommand \bibinfo  [0]{\@secondoftwo}%
\providecommand \bibfield  [0]{\@secondoftwo}%
\providecommand \translation [1]{[#1]}%
\providecommand \BibitemOpen [0]{}%
\providecommand \bibitemStop [0]{}%
\providecommand \bibitemNoStop [0]{.\EOS\space}%
\providecommand \EOS [0]{\spacefactor3000\relax}%
\providecommand \BibitemShut  [1]{\csname bibitem#1\endcsname}%
\let\auto@bib@innerbib\@empty
\bibitem [{\citenamefont {Schirmer}\ and\ \citenamefont
  {Langbein}(2015)}]{PhysRevA.91.022125}%
  \BibitemOpen
  \bibfield  {author} {\bibinfo {author} {\bibfnamefont {S.~G.}\ \bibnamefont
  {Schirmer}}\ and\ \bibinfo {author} {\bibfnamefont {F.~C.}\ \bibnamefont
  {Langbein}},\ }\href {\doibase 10.1103/PhysRevA.91.022125} {\bibfield
  {journal} {\bibinfo  {journal} {Phys. Rev. A}\ }\textbf {\bibinfo {volume}
  {91}},\ \bibinfo {pages} {022125} (\bibinfo {year} {2015})}\BibitemShut
  {NoStop}%
\bibitem [{\citenamefont {Carvalho}\ \emph {et~al.}(2021)\citenamefont
  {Carvalho}, \citenamefont {Ball}, \citenamefont {Biercuk}, \citenamefont
  {Hush},\ and\ \citenamefont {Thomsen}}]{PhysRevApplied.15.064054}%
  \BibitemOpen
  \bibfield  {author} {\bibinfo {author} {\bibfnamefont {A.~R.~R.}\
  \bibnamefont {Carvalho}}, \bibinfo {author} {\bibfnamefont {H.}~\bibnamefont
  {Ball}}, \bibinfo {author} {\bibfnamefont {M.~J.}\ \bibnamefont {Biercuk}},
  \bibinfo {author} {\bibfnamefont {M.~R.}\ \bibnamefont {Hush}}, \ and\
  \bibinfo {author} {\bibfnamefont {F.}~\bibnamefont {Thomsen}},\ }\href
  {\doibase 10.1103/PhysRevApplied.15.064054} {\bibfield  {journal} {\bibinfo
  {journal} {Phys. Rev. Applied}\ }\textbf {\bibinfo {volume} {15}},\ \bibinfo
  {pages} {064054} (\bibinfo {year} {2021})}\BibitemShut {NoStop}%
\bibitem [{\citenamefont {Mart{\'{\i}}nez-Garc{\'{\i}}a}\ \emph
  {et~al.}(2019)\citenamefont {Mart{\'{\i}}nez-Garc{\'{\i}}a}, \citenamefont
  {Vodola},\ and\ \citenamefont {M{\"u}ller}}]{Mart_nez_Garc_a_2019}%
  \BibitemOpen
  \bibfield  {author} {\bibinfo {author} {\bibfnamefont {F.}~\bibnamefont
  {Mart{\'{\i}}nez-Garc{\'{\i}}a}}, \bibinfo {author} {\bibfnamefont
  {D.}~\bibnamefont {Vodola}}, \ and\ \bibinfo {author} {\bibfnamefont
  {M.}~\bibnamefont {M{\"u}ller}},\ }\href {\doibase 10.1088/1367-2630/ab5c51}
  {\bibfield  {journal} {\bibinfo  {journal} {New Journal of Physics}\ }\textbf
  {\bibinfo {volume} {21}},\ \bibinfo {pages} {123027} (\bibinfo {year}
  {2019})}\BibitemShut {NoStop}%
\bibitem [{\citenamefont {Aliferis}\ \emph {et~al.}(2008)\citenamefont
  {Aliferis}, \citenamefont {Gottesman},\ and\ \citenamefont
  {Preskill}}]{10.5555/2011763.2011764}%
  \BibitemOpen
  \bibfield  {author} {\bibinfo {author} {\bibfnamefont {P.}~\bibnamefont
  {Aliferis}}, \bibinfo {author} {\bibfnamefont {D.}~\bibnamefont {Gottesman}},
  \ and\ \bibinfo {author} {\bibfnamefont {J.}~\bibnamefont {Preskill}},\
  }\href@noop {} {\bibfield  {journal} {\bibinfo  {journal} {Quantum Info.
  Comput.}\ }\textbf {\bibinfo {volume} {8}},\ \bibinfo {pages} {181–244}
  (\bibinfo {year} {2008})}\BibitemShut {NoStop}%
\bibitem [{\citenamefont {Kjaergaard}\ \emph {et~al.}(2020)\citenamefont
  {Kjaergaard}, \citenamefont {Schwartz}, \citenamefont {Braum\"{u}ller},
  \citenamefont {Krantz}, \citenamefont {Wang}, \citenamefont {Gustavsson},\
  and\ \citenamefont {Oliver}}]{doi:10.1146_annurev-conmatphys-031119-050605}%
  \BibitemOpen
  \bibfield  {author} {\bibinfo {author} {\bibfnamefont {M.}~\bibnamefont
  {Kjaergaard}}, \bibinfo {author} {\bibfnamefont {M.~E.}\ \bibnamefont
  {Schwartz}}, \bibinfo {author} {\bibfnamefont {J.}~\bibnamefont
  {Braum\"{u}ller}}, \bibinfo {author} {\bibfnamefont {P.}~\bibnamefont
  {Krantz}}, \bibinfo {author} {\bibfnamefont {J.~I.-J.}\ \bibnamefont {Wang}},
  \bibinfo {author} {\bibfnamefont {S.}~\bibnamefont {Gustavsson}}, \ and\
  \bibinfo {author} {\bibfnamefont {W.~D.}\ \bibnamefont {Oliver}},\ }\href
  {\doibase 10.1146/annurev-conmatphys-031119-050605} {\bibfield  {journal}
  {\bibinfo  {journal} {Annual Review of Condensed Matter Physics}\ }\textbf
  {\bibinfo {volume} {11}},\ \bibinfo {pages} {369} (\bibinfo {year} {2020})},\
  \Eprint
  {http://arxiv.org/abs/https://doi.org/10.1146/annurev-conmatphys-031119-050605}
  {https://doi.org/10.1146/annurev-conmatphys-031119-050605} \BibitemShut
  {NoStop}%
\bibitem [{\citenamefont {Nielsen}\ \emph {et~al.}(2021)\citenamefont
  {Nielsen}, \citenamefont {Gamble}, \citenamefont {Rudinger}, \citenamefont
  {Scholten}, \citenamefont {Young},\ and\ \citenamefont
  {Blume-Kohout}}]{Nielsen2021gatesettomography}%
  \BibitemOpen
  \bibfield  {author} {\bibinfo {author} {\bibfnamefont {E.}~\bibnamefont
  {Nielsen}}, \bibinfo {author} {\bibfnamefont {J.~K.}\ \bibnamefont {Gamble}},
  \bibinfo {author} {\bibfnamefont {K.}~\bibnamefont {Rudinger}}, \bibinfo
  {author} {\bibfnamefont {T.}~\bibnamefont {Scholten}}, \bibinfo {author}
  {\bibfnamefont {K.}~\bibnamefont {Young}}, \ and\ \bibinfo {author}
  {\bibfnamefont {R.}~\bibnamefont {Blume-Kohout}},\ }\href {\doibase
  10.22331/q-2021-10-05-557} {\bibfield  {journal} {\bibinfo  {journal}
  {{Quantum}}\ }\textbf {\bibinfo {volume} {5}},\ \bibinfo {pages} {557}
  (\bibinfo {year} {2021})}\BibitemShut {NoStop}%
\bibitem [{\citenamefont {Takita}\ \emph {et~al.}(2017)\citenamefont {Takita},
  \citenamefont {Cross}, \citenamefont {C\'orcoles}, \citenamefont {Chow},\
  and\ \citenamefont {Gambetta}}]{PhysRevLett.119.180501}%
  \BibitemOpen
  \bibfield  {author} {\bibinfo {author} {\bibfnamefont {M.}~\bibnamefont
  {Takita}}, \bibinfo {author} {\bibfnamefont {A.~W.}\ \bibnamefont {Cross}},
  \bibinfo {author} {\bibfnamefont {A.~D.}\ \bibnamefont {C\'orcoles}},
  \bibinfo {author} {\bibfnamefont {J.~M.}\ \bibnamefont {Chow}}, \ and\
  \bibinfo {author} {\bibfnamefont {J.~M.}\ \bibnamefont {Gambetta}},\ }\href
  {\doibase 10.1103/PhysRevLett.119.180501} {\bibfield  {journal} {\bibinfo
  {journal} {Phys. Rev. Lett.}\ }\textbf {\bibinfo {volume} {119}},\ \bibinfo
  {pages} {180501} (\bibinfo {year} {2017})}\BibitemShut {NoStop}%
\bibitem [{\citenamefont {Chen}\ \emph {et~al.}(2021)\citenamefont {Chen},
  \citenamefont {Satzinger}, \citenamefont {Atalaya}, \citenamefont {Korotkov}
  \emph {et~al.}}]{Google2021}%
  \BibitemOpen
  \bibfield  {author} {\bibinfo {author} {\bibfnamefont {Z.}~\bibnamefont
  {Chen}}, \bibinfo {author} {\bibfnamefont {K.~J.}\ \bibnamefont {Satzinger}},
  \bibinfo {author} {\bibfnamefont {J.}~\bibnamefont {Atalaya}}, \bibinfo
  {author} {\bibfnamefont {A.~N.}\ \bibnamefont {Korotkov}},  \emph {et~al.},\
  }\href {\doibase 10.1038/s41586-021-03588-y} {\bibfield  {journal} {\bibinfo
  {journal} {Nature}\ }\textbf {\bibinfo {volume} {595}},\ \bibinfo {pages}
  {383} (\bibinfo {year} {2021})}\BibitemShut {NoStop}%
\bibitem [{\citenamefont {Pino}\ \emph {et~al.}(2021)\citenamefont {Pino},
  \citenamefont {Dreiling}, \citenamefont {Figgatt}, \citenamefont {Gaebler},
  \citenamefont {Moses}, \citenamefont {Allman}, \citenamefont {Baldwin},
  \citenamefont {Foss-Feig}, \citenamefont {Hayes}, \citenamefont {Mayer},
  \citenamefont {Ryan-Anderson},\ and\ \citenamefont
  {Neyenhuis}}]{Pino:2021tf}%
  \BibitemOpen
  \bibfield  {author} {\bibinfo {author} {\bibfnamefont {J.~M.}\ \bibnamefont
  {Pino}}, \bibinfo {author} {\bibfnamefont {J.~M.}\ \bibnamefont {Dreiling}},
  \bibinfo {author} {\bibfnamefont {C.}~\bibnamefont {Figgatt}}, \bibinfo
  {author} {\bibfnamefont {J.~P.}\ \bibnamefont {Gaebler}}, \bibinfo {author}
  {\bibfnamefont {S.~A.}\ \bibnamefont {Moses}}, \bibinfo {author}
  {\bibfnamefont {M.~S.}\ \bibnamefont {Allman}}, \bibinfo {author}
  {\bibfnamefont {C.~H.}\ \bibnamefont {Baldwin}}, \bibinfo {author}
  {\bibfnamefont {M.}~\bibnamefont {Foss-Feig}}, \bibinfo {author}
  {\bibfnamefont {D.}~\bibnamefont {Hayes}}, \bibinfo {author} {\bibfnamefont
  {K.}~\bibnamefont {Mayer}}, \bibinfo {author} {\bibfnamefont
  {C.}~\bibnamefont {Ryan-Anderson}}, \ and\ \bibinfo {author} {\bibfnamefont
  {B.}~\bibnamefont {Neyenhuis}},\ }\href {\doibase 10.1038/s41586-021-03318-4}
  {\bibfield  {journal} {\bibinfo  {journal} {Nature}\ }\textbf {\bibinfo
  {volume} {592}},\ \bibinfo {pages} {209} (\bibinfo {year}
  {2021})}\BibitemShut {NoStop}%
\bibitem [{\citenamefont {Postler}\ \emph {et~al.}(2022)\citenamefont
  {Postler}, \citenamefont {Heussen}, \citenamefont {Pogorelov}, \citenamefont
  {Rispler}, \citenamefont {Feldker}, \citenamefont {Meth}, \citenamefont
  {Marciniak}, \citenamefont {Stricker}, \citenamefont {Ringbauer},
  \citenamefont {Blatt}, \citenamefont {Schindler}, \citenamefont
  {M{\"u}ller},\ and\ \citenamefont {Monz}}]{Postler:2022uq}%
  \BibitemOpen
  \bibfield  {author} {\bibinfo {author} {\bibfnamefont {L.}~\bibnamefont
  {Postler}}, \bibinfo {author} {\bibfnamefont {S.}~\bibnamefont {Heussen}},
  \bibinfo {author} {\bibfnamefont {I.}~\bibnamefont {Pogorelov}}, \bibinfo
  {author} {\bibfnamefont {M.}~\bibnamefont {Rispler}}, \bibinfo {author}
  {\bibfnamefont {T.}~\bibnamefont {Feldker}}, \bibinfo {author} {\bibfnamefont
  {M.}~\bibnamefont {Meth}}, \bibinfo {author} {\bibfnamefont {C.~D.}\
  \bibnamefont {Marciniak}}, \bibinfo {author} {\bibfnamefont {R.}~\bibnamefont
  {Stricker}}, \bibinfo {author} {\bibfnamefont {M.}~\bibnamefont {Ringbauer}},
  \bibinfo {author} {\bibfnamefont {R.}~\bibnamefont {Blatt}}, \bibinfo
  {author} {\bibfnamefont {P.}~\bibnamefont {Schindler}}, \bibinfo {author}
  {\bibfnamefont {M.}~\bibnamefont {M{\"u}ller}}, \ and\ \bibinfo {author}
  {\bibfnamefont {T.}~\bibnamefont {Monz}},\ }\href {\doibase
  10.1038/s41586-022-04721-1} {\bibfield  {journal} {\bibinfo  {journal}
  {Nature}\ }\textbf {\bibinfo {volume} {605}},\ \bibinfo {pages} {675}
  (\bibinfo {year} {2022})}\BibitemShut {NoStop}%
\bibitem [{\citenamefont {Krinner}\ \emph {et~al.}(2022)\citenamefont
  {Krinner}, \citenamefont {Lacroix}, \citenamefont {Remm}, \citenamefont
  {Di~Paolo}, \citenamefont {Genois}, \citenamefont {Leroux}, \citenamefont
  {Hellings}, \citenamefont {Lazar}, \citenamefont {Swiadek}, \citenamefont
  {Herrmann}, \citenamefont {Norris}, \citenamefont {Andersen}, \citenamefont
  {M{\"u}ller}, \citenamefont {Blais}, \citenamefont {Eichler},\ and\
  \citenamefont {Wallraff}}]{ETHsurface}%
  \BibitemOpen
  \bibfield  {author} {\bibinfo {author} {\bibfnamefont {S.}~\bibnamefont
  {Krinner}}, \bibinfo {author} {\bibfnamefont {N.}~\bibnamefont {Lacroix}},
  \bibinfo {author} {\bibfnamefont {A.}~\bibnamefont {Remm}}, \bibinfo {author}
  {\bibfnamefont {A.}~\bibnamefont {Di~Paolo}}, \bibinfo {author}
  {\bibfnamefont {E.}~\bibnamefont {Genois}}, \bibinfo {author} {\bibfnamefont
  {C.}~\bibnamefont {Leroux}}, \bibinfo {author} {\bibfnamefont
  {C.}~\bibnamefont {Hellings}}, \bibinfo {author} {\bibfnamefont
  {S.}~\bibnamefont {Lazar}}, \bibinfo {author} {\bibfnamefont
  {F.}~\bibnamefont {Swiadek}}, \bibinfo {author} {\bibfnamefont
  {J.}~\bibnamefont {Herrmann}}, \bibinfo {author} {\bibfnamefont {G.~J.}\
  \bibnamefont {Norris}}, \bibinfo {author} {\bibfnamefont {C.~K.}\
  \bibnamefont {Andersen}}, \bibinfo {author} {\bibfnamefont {M.}~\bibnamefont
  {M{\"u}ller}}, \bibinfo {author} {\bibfnamefont {A.}~\bibnamefont {Blais}},
  \bibinfo {author} {\bibfnamefont {C.}~\bibnamefont {Eichler}}, \ and\
  \bibinfo {author} {\bibfnamefont {A.}~\bibnamefont {Wallraff}},\ }\href
  {\doibase 10.1038/s41586-022-04566-8} {\bibfield  {journal} {\bibinfo
  {journal} {Nature}\ }\textbf {\bibinfo {volume} {605}},\ \bibinfo {pages}
  {669} (\bibinfo {year} {2022})}\BibitemShut {NoStop}%
\bibitem [{\citenamefont {Foster}\ \emph {et~al.}(2019)\citenamefont {Foster},
  \citenamefont {Jankowiak}, \citenamefont {Bingham}, \citenamefont {Horsfall},
  \citenamefont {Teh}, \citenamefont {Rainforth},\ and\ \citenamefont
  {Goodman}}]{NEURIPS2019_d55cbf21}%
  \BibitemOpen
  \bibfield  {author} {\bibinfo {author} {\bibfnamefont {A.}~\bibnamefont
  {Foster}}, \bibinfo {author} {\bibfnamefont {M.}~\bibnamefont {Jankowiak}},
  \bibinfo {author} {\bibfnamefont {E.}~\bibnamefont {Bingham}}, \bibinfo
  {author} {\bibfnamefont {P.}~\bibnamefont {Horsfall}}, \bibinfo {author}
  {\bibfnamefont {Y.~W.}\ \bibnamefont {Teh}}, \bibinfo {author} {\bibfnamefont
  {T.}~\bibnamefont {Rainforth}}, \ and\ \bibinfo {author} {\bibfnamefont
  {N.}~\bibnamefont {Goodman}},\ }in\ \href
  {https://proceedings.neurips.cc/paper/2019/file/d55cbf210f175f4a37916eafe6c04f0d-Paper.pdf}
  {\emph {\bibinfo {booktitle} {Advances in Neural Information Processing
  Systems}}},\ Vol.~\bibinfo {volume} {32},\ \bibinfo {editor} {edited by\
  \bibinfo {editor} {\bibfnamefont {H.}~\bibnamefont {Wallach}}, \bibinfo
  {editor} {\bibfnamefont {H.}~\bibnamefont {Larochelle}}, \bibinfo {editor}
  {\bibfnamefont {A.}~\bibnamefont {Beygelzimer}}, \bibinfo {editor}
  {\bibfnamefont {F.}~\bibnamefont {d\textquotesingle Alch\'{e}-Buc}}, \bibinfo
  {editor} {\bibfnamefont {E.}~\bibnamefont {Fox}}, \ and\ \bibinfo {editor}
  {\bibfnamefont {R.}~\bibnamefont {Garnett}}}\ (\bibinfo  {publisher} {Curran
  Associates, Inc.},\ \bibinfo {year} {2019})\BibitemShut {NoStop}%
\bibitem [{\citenamefont {Foster}\ \emph {et~al.}(2020)\citenamefont {Foster},
  \citenamefont {Jankowiak}, \citenamefont {O'Meara}, \citenamefont {Teh},\
  and\ \citenamefont {Rainforth}}]{pmlr-v108-foster20a}%
  \BibitemOpen
  \bibfield  {author} {\bibinfo {author} {\bibfnamefont {A.}~\bibnamefont
  {Foster}}, \bibinfo {author} {\bibfnamefont {M.}~\bibnamefont {Jankowiak}},
  \bibinfo {author} {\bibfnamefont {M.}~\bibnamefont {O'Meara}}, \bibinfo
  {author} {\bibfnamefont {Y.~W.}\ \bibnamefont {Teh}}, \ and\ \bibinfo
  {author} {\bibfnamefont {T.}~\bibnamefont {Rainforth}},\ }in\ \href
  {https://proceedings.mlr.press/v108/foster20a.html} {\emph {\bibinfo
  {booktitle} {Proceedings of the Twenty Third International Conference on
  Artificial Intelligence and Statistics}}},\ \bibinfo {series} {Proceedings of
  Machine Learning Research}, Vol.\ \bibinfo {volume} {108},\ \bibinfo {editor}
  {edited by\ \bibinfo {editor} {\bibfnamefont {S.}~\bibnamefont {Chiappa}}\
  and\ \bibinfo {editor} {\bibfnamefont {R.}~\bibnamefont {Calandra}}}\
  (\bibinfo  {publisher} {PMLR},\ \bibinfo {year} {2020})\ pp.\ \bibinfo
  {pages} {2959--2969}\BibitemShut {NoStop}%
\bibitem [{\citenamefont {Baum}\ \emph {et~al.}(2021)\citenamefont {Baum},
  \citenamefont {Amico}, \citenamefont {Howell}, \citenamefont {Hush},
  \citenamefont {Liuzzi}, \citenamefont {Mundada}, \citenamefont {Merkh},
  \citenamefont {Carvalho},\ and\ \citenamefont
  {Biercuk}}]{PRXQuantum.2.040324}%
  \BibitemOpen
  \bibfield  {author} {\bibinfo {author} {\bibfnamefont {Y.}~\bibnamefont
  {Baum}}, \bibinfo {author} {\bibfnamefont {M.}~\bibnamefont {Amico}},
  \bibinfo {author} {\bibfnamefont {S.}~\bibnamefont {Howell}}, \bibinfo
  {author} {\bibfnamefont {M.}~\bibnamefont {Hush}}, \bibinfo {author}
  {\bibfnamefont {M.}~\bibnamefont {Liuzzi}}, \bibinfo {author} {\bibfnamefont
  {P.}~\bibnamefont {Mundada}}, \bibinfo {author} {\bibfnamefont
  {T.}~\bibnamefont {Merkh}}, \bibinfo {author} {\bibfnamefont {A.~R.}\
  \bibnamefont {Carvalho}}, \ and\ \bibinfo {author} {\bibfnamefont {M.~J.}\
  \bibnamefont {Biercuk}},\ }\href {\doibase 10.1103/PRXQuantum.2.040324}
  {\bibfield  {journal} {\bibinfo  {journal} {PRX Quantum}\ }\textbf {\bibinfo
  {volume} {2}},\ \bibinfo {pages} {040324} (\bibinfo {year}
  {2021})}\BibitemShut {NoStop}%
\bibitem [{\citenamefont {Arulampalam}\ \emph {et~al.}(2002)\citenamefont
  {Arulampalam}, \citenamefont {Maskell}, \citenamefont {Gordon},\ and\
  \citenamefont {Clapp}}]{978374}%
  \BibitemOpen
  \bibfield  {author} {\bibinfo {author} {\bibfnamefont {M.}~\bibnamefont
  {Arulampalam}}, \bibinfo {author} {\bibfnamefont {S.}~\bibnamefont
  {Maskell}}, \bibinfo {author} {\bibfnamefont {N.}~\bibnamefont {Gordon}}, \
  and\ \bibinfo {author} {\bibfnamefont {T.}~\bibnamefont {Clapp}},\ }\href
  {\doibase 10.1109/78.978374} {\bibfield  {journal} {\bibinfo  {journal} {IEEE
  Transactions on Signal Processing}\ }\textbf {\bibinfo {volume} {50}},\
  \bibinfo {pages} {174} (\bibinfo {year} {2002})}\BibitemShut {NoStop}%
\bibitem [{\citenamefont {Gerster}\ \emph {et~al.}(2022)\citenamefont
  {Gerster}, \citenamefont {Mart\'{\i}nez-Garc\'{\i}a}, \citenamefont {Hrmo},
  \citenamefont {van Mourik}, \citenamefont {Wilhelm}, \citenamefont {Vodola},
  \citenamefont {M\"uller}, \citenamefont {Blatt}, \citenamefont {Schindler},\
  and\ \citenamefont {Monz}}]{PRXQuantum.3.020350}%
  \BibitemOpen
  \bibfield  {author} {\bibinfo {author} {\bibfnamefont {L.}~\bibnamefont
  {Gerster}}, \bibinfo {author} {\bibfnamefont {F.}~\bibnamefont
  {Mart\'{\i}nez-Garc\'{\i}a}}, \bibinfo {author} {\bibfnamefont
  {P.}~\bibnamefont {Hrmo}}, \bibinfo {author} {\bibfnamefont {M.~W.}\
  \bibnamefont {van Mourik}}, \bibinfo {author} {\bibfnamefont
  {B.}~\bibnamefont {Wilhelm}}, \bibinfo {author} {\bibfnamefont
  {D.}~\bibnamefont {Vodola}}, \bibinfo {author} {\bibfnamefont
  {M.}~\bibnamefont {M\"uller}}, \bibinfo {author} {\bibfnamefont
  {R.}~\bibnamefont {Blatt}}, \bibinfo {author} {\bibfnamefont
  {P.}~\bibnamefont {Schindler}}, \ and\ \bibinfo {author} {\bibfnamefont
  {T.}~\bibnamefont {Monz}},\ }\href {\doibase 10.1103/PRXQuantum.3.020350}
  {\bibfield  {journal} {\bibinfo  {journal} {PRX Quantum}\ }\textbf {\bibinfo
  {volume} {3}},\ \bibinfo {pages} {020350} (\bibinfo {year}
  {2022})}\BibitemShut {NoStop}%
\bibitem [{\citenamefont {Kok}\ and\ \citenamefont
  {Lovett}(2010)}]{kok_lovett_2010}%
  \BibitemOpen
  \bibfield  {author} {\bibinfo {author} {\bibfnamefont {P.}~\bibnamefont
  {Kok}}\ and\ \bibinfo {author} {\bibfnamefont {B.~W.}\ \bibnamefont
  {Lovett}},\ }\href {\doibase 10.1017/CBO9781139193658} {\emph {\bibinfo
  {title} {Introduction to Optical Quantum Information Processing}}}\ (\bibinfo
   {publisher} {Cambridge University Press},\ \bibinfo {year}
  {2010})\BibitemShut {NoStop}%
\bibitem [{\citenamefont {Kiilerich}\ and\ \citenamefont
  {M\o{}lmer}(2015)}]{PhysRevA.92.032124}%
  \BibitemOpen
  \bibfield  {author} {\bibinfo {author} {\bibfnamefont {A.~H.}\ \bibnamefont
  {Kiilerich}}\ and\ \bibinfo {author} {\bibfnamefont {K.}~\bibnamefont
  {M\o{}lmer}},\ }\href {\doibase 10.1103/PhysRevA.92.032124} {\bibfield
  {journal} {\bibinfo  {journal} {Phys. Rev. A}\ }\textbf {\bibinfo {volume}
  {92}},\ \bibinfo {pages} {032124} (\bibinfo {year} {2015})}\BibitemShut
  {NoStop}%
\bibitem [{\citenamefont {Draper}\ and\ \citenamefont
  {Guttman}(1995)}]{10.2307_2348711}%
  \BibitemOpen
  \bibfield  {author} {\bibinfo {author} {\bibfnamefont {N.~R.}\ \bibnamefont
  {Draper}}\ and\ \bibinfo {author} {\bibfnamefont {I.}~\bibnamefont
  {Guttman}},\ }\href {http://www.jstor.org/stable/2348711} {\bibfield
  {journal} {\bibinfo  {journal} {Journal of the Royal Statistical Society.
  Series D (The Statistician)}\ }\textbf {\bibinfo {volume} {44}},\ \bibinfo
  {pages} {399} (\bibinfo {year} {1995})}\BibitemShut {NoStop}%
\bibitem [{\citenamefont {Tokdar}\ and\ \citenamefont
  {Kass}(2010)}]{tokdar2010importance}%
  \BibitemOpen
  \bibfield  {author} {\bibinfo {author} {\bibfnamefont {S.~T.}\ \bibnamefont
  {Tokdar}}\ and\ \bibinfo {author} {\bibfnamefont {R.~E.}\ \bibnamefont
  {Kass}},\ }\href@noop {} {\bibfield  {journal} {\bibinfo  {journal} {Wiley
  Interdisciplinary Reviews: Computational Statistics}\ }\textbf {\bibinfo
  {volume} {2}},\ \bibinfo {pages} {54} (\bibinfo {year} {2010})}\BibitemShut
  {NoStop}%
\bibitem [{\citenamefont {Ball}\ \emph {et~al.}(2021)\citenamefont {Ball},
  \citenamefont {Biercuk}, \citenamefont {Carvalho}, \citenamefont {Chen},
  \citenamefont {Hush}, \citenamefont {Castro}, \citenamefont {Li},
  \citenamefont {Liebermann}, \citenamefont {Slatyer}, \citenamefont {Edmunds},
  \citenamefont {Frey}, \citenamefont {Hempel},\ and\ \citenamefont
  {Milne}}]{Ball_2021}%
  \BibitemOpen
  \bibfield  {author} {\bibinfo {author} {\bibfnamefont {H.}~\bibnamefont
  {Ball}}, \bibinfo {author} {\bibfnamefont {M.~J.}\ \bibnamefont {Biercuk}},
  \bibinfo {author} {\bibfnamefont {A.~R.~R.}\ \bibnamefont {Carvalho}},
  \bibinfo {author} {\bibfnamefont {J.}~\bibnamefont {Chen}}, \bibinfo {author}
  {\bibfnamefont {M.}~\bibnamefont {Hush}}, \bibinfo {author} {\bibfnamefont
  {L.~A.~D.}\ \bibnamefont {Castro}}, \bibinfo {author} {\bibfnamefont
  {L.}~\bibnamefont {Li}}, \bibinfo {author} {\bibfnamefont {P.~J.}\
  \bibnamefont {Liebermann}}, \bibinfo {author} {\bibfnamefont {H.~J.}\
  \bibnamefont {Slatyer}}, \bibinfo {author} {\bibfnamefont {C.}~\bibnamefont
  {Edmunds}}, \bibinfo {author} {\bibfnamefont {V.}~\bibnamefont {Frey}},
  \bibinfo {author} {\bibfnamefont {C.}~\bibnamefont {Hempel}}, \ and\ \bibinfo
  {author} {\bibfnamefont {A.}~\bibnamefont {Milne}},\ }\href {\doibase
  10.1088/2058-9565/abdca6} {\bibfield  {journal} {\bibinfo  {journal} {Quantum
  Science and Technology}\ }\textbf {\bibinfo {volume} {6}},\ \bibinfo {pages}
  {044011} (\bibinfo {year} {2021})}\BibitemShut {NoStop}%
\bibitem [{\citenamefont {Q-CTRL}(2022)}]{boulder_opal2}%
  \BibitemOpen
  \bibfield  {author} {\bibinfo {author} {\bibnamefont {Q-CTRL}},\ }\href@noop
  {} {\enquote {\bibinfo {title} {Boulder {O}pal},}\ }\bibinfo {howpublished}
  {https://q-ctrl.com/boulder-opal} (\bibinfo {year} {2022}),\ \bibinfo {note}
  {[Online]}\BibitemShut {NoStop}%
\bibitem [{\citenamefont {Abadi}\ \emph {et~al.}(2015)\citenamefont {Abadi},
  \citenamefont {Agarwal}, \citenamefont {Barham}, \citenamefont {Brevdo},
  \citenamefont {Chen} \emph {et~al.}}]{tensorflow2015-whitepaper}%
  \BibitemOpen
  \bibfield  {author} {\bibinfo {author} {\bibfnamefont {M.}~\bibnamefont
  {Abadi}}, \bibinfo {author} {\bibfnamefont {A.}~\bibnamefont {Agarwal}},
  \bibinfo {author} {\bibfnamefont {P.}~\bibnamefont {Barham}}, \bibinfo
  {author} {\bibfnamefont {E.}~\bibnamefont {Brevdo}}, \bibinfo {author}
  {\bibfnamefont {Z.}~\bibnamefont {Chen}},  \emph {et~al.},\ }\href
  {https://www.tensorflow.org/} {\enquote {\bibinfo {title} {{TensorFlow}:
  Large-scale machine learning on heterogeneous systems},}\ } (\bibinfo {year}
  {2015}),\ \bibinfo {note} {software available from
  tensorflow.org}\BibitemShut {NoStop}%
\bibitem [{\citenamefont {Stuhlm\"{u}ller}\ \emph {et~al.}(2013)\citenamefont
  {Stuhlm\"{u}ller}, \citenamefont {Taylor},\ and\ \citenamefont
  {Goodman}}]{NIPS2013_7f53f8c6}%
  \BibitemOpen
  \bibfield  {author} {\bibinfo {author} {\bibfnamefont {A.}~\bibnamefont
  {Stuhlm\"{u}ller}}, \bibinfo {author} {\bibfnamefont {J.}~\bibnamefont
  {Taylor}}, \ and\ \bibinfo {author} {\bibfnamefont {N.}~\bibnamefont
  {Goodman}},\ }in\ \href
  {https://proceedings.neurips.cc/paper/2013/file/7f53f8c6c730af6aeb52e66eb74d8507-Paper.pdf}
  {\emph {\bibinfo {booktitle} {Advances in Neural Information Processing
  Systems}}},\ Vol.~\bibinfo {volume} {26},\ \bibinfo {editor} {edited by\
  \bibinfo {editor} {\bibfnamefont {C.}~\bibnamefont {Burges}}, \bibinfo
  {editor} {\bibfnamefont {L.}~\bibnamefont {Bottou}}, \bibinfo {editor}
  {\bibfnamefont {M.}~\bibnamefont {Welling}}, \bibinfo {editor} {\bibfnamefont
  {Z.}~\bibnamefont {Ghahramani}}, \ and\ \bibinfo {editor} {\bibfnamefont
  {K.}~\bibnamefont {Weinberger}}}\ (\bibinfo {year} {2013})\BibitemShut
  {NoStop}%
\bibitem [{\citenamefont {Milne}\ \emph {et~al.}(2020)\citenamefont {Milne},
  \citenamefont {Edmunds}, \citenamefont {Hempel}, \citenamefont {Roy},
  \citenamefont {Mavadia},\ and\ \citenamefont {Biercuk}}]{Milne2020}%
  \BibitemOpen
  \bibfield  {author} {\bibinfo {author} {\bibfnamefont {A.~R.}\ \bibnamefont
  {Milne}}, \bibinfo {author} {\bibfnamefont {C.~L.}\ \bibnamefont {Edmunds}},
  \bibinfo {author} {\bibfnamefont {C.}~\bibnamefont {Hempel}}, \bibinfo
  {author} {\bibfnamefont {F.}~\bibnamefont {Roy}}, \bibinfo {author}
  {\bibfnamefont {S.}~\bibnamefont {Mavadia}}, \ and\ \bibinfo {author}
  {\bibfnamefont {M.~J.}\ \bibnamefont {Biercuk}},\ }\href {\doibase
  10.1103/physrevapplied.13.024022} {\bibfield  {journal} {\bibinfo  {journal}
  {Physical Review Applied}\ }\textbf {\bibinfo {volume} {13}} (\bibinfo {year}
  {2020}),\ 10.1103/physrevapplied.13.024022}\BibitemShut {NoStop}%
\bibitem [{\citenamefont {Olmschenk}\ \emph {et~al.}(2007)\citenamefont
  {Olmschenk}, \citenamefont {Younge}, \citenamefont {Moehring}, \citenamefont
  {Matsukevich}, \citenamefont {Maunz},\ and\ \citenamefont
  {Monroe}}]{Olmschenk2007}%
  \BibitemOpen
  \bibfield  {author} {\bibinfo {author} {\bibfnamefont {S.}~\bibnamefont
  {Olmschenk}}, \bibinfo {author} {\bibfnamefont {K.~C.}\ \bibnamefont
  {Younge}}, \bibinfo {author} {\bibfnamefont {D.~L.}\ \bibnamefont
  {Moehring}}, \bibinfo {author} {\bibfnamefont {D.~N.}\ \bibnamefont
  {Matsukevich}}, \bibinfo {author} {\bibfnamefont {P.}~\bibnamefont {Maunz}},
  \ and\ \bibinfo {author} {\bibfnamefont {C.}~\bibnamefont {Monroe}},\ }\href
  {\doibase 10.1103/physreva.76.052314} {\bibfield  {journal} {\bibinfo
  {journal} {Physical Review A}\ }\textbf {\bibinfo {volume} {76}} (\bibinfo
  {year} {2007}),\ 10.1103/physreva.76.052314}\BibitemShut {NoStop}%
\bibitem [{\citenamefont {Peik}\ \emph {et~al.}(2005)\citenamefont {Peik},
  \citenamefont {Schneider},\ and\ \citenamefont {Tamm}}]{Peik2005}%
  \BibitemOpen
  \bibfield  {author} {\bibinfo {author} {\bibfnamefont {E.}~\bibnamefont
  {Peik}}, \bibinfo {author} {\bibfnamefont {T.}~\bibnamefont {Schneider}}, \
  and\ \bibinfo {author} {\bibfnamefont {C.}~\bibnamefont {Tamm}},\ }\href
  {\doibase 10.1088/0953-4075/39/1/012} {\bibfield  {journal} {\bibinfo
  {journal} {Journal of Physics B: Atomic, Molecular and Optical Physics}\
  }\textbf {\bibinfo {volume} {39}},\ \bibinfo {pages} {145} (\bibinfo {year}
  {2005})}\BibitemShut {NoStop}%
\bibitem [{\citenamefont {Chow}\ \emph {et~al.}(2011)\citenamefont {Chow},
  \citenamefont {C\'orcoles}, \citenamefont {Gambetta}, \citenamefont
  {Rigetti}, \citenamefont {Johnson}, \citenamefont {Smolin}, \citenamefont
  {Rozen}, \citenamefont {Keefe}, \citenamefont {Rothwell}, \citenamefont
  {Ketchen},\ and\ \citenamefont {Steffen}}]{PhysRevLett.107.080502}%
  \BibitemOpen
  \bibfield  {author} {\bibinfo {author} {\bibfnamefont {J.~M.}\ \bibnamefont
  {Chow}}, \bibinfo {author} {\bibfnamefont {A.~D.}\ \bibnamefont
  {C\'orcoles}}, \bibinfo {author} {\bibfnamefont {J.~M.}\ \bibnamefont
  {Gambetta}}, \bibinfo {author} {\bibfnamefont {C.}~\bibnamefont {Rigetti}},
  \bibinfo {author} {\bibfnamefont {B.~R.}\ \bibnamefont {Johnson}}, \bibinfo
  {author} {\bibfnamefont {J.~A.}\ \bibnamefont {Smolin}}, \bibinfo {author}
  {\bibfnamefont {J.~R.}\ \bibnamefont {Rozen}}, \bibinfo {author}
  {\bibfnamefont {G.~A.}\ \bibnamefont {Keefe}}, \bibinfo {author}
  {\bibfnamefont {M.~B.}\ \bibnamefont {Rothwell}}, \bibinfo {author}
  {\bibfnamefont {M.~B.}\ \bibnamefont {Ketchen}}, \ and\ \bibinfo {author}
  {\bibfnamefont {M.}~\bibnamefont {Steffen}},\ }\href {\doibase
  10.1103/PhysRevLett.107.080502} {\bibfield  {journal} {\bibinfo  {journal}
  {Phys. Rev. Lett.}\ }\textbf {\bibinfo {volume} {107}},\ \bibinfo {pages}
  {080502} (\bibinfo {year} {2011})}\BibitemShut {NoStop}%
\bibitem [{\citenamefont {Yang}\ \emph {et~al.}(2018)\citenamefont {Yang},
  \citenamefont {Gong},\ and\ \citenamefont {Cui}}]{PhysRevA.97.012119}%
  \BibitemOpen
  \bibfield  {author} {\bibinfo {author} {\bibfnamefont {Y.}~\bibnamefont
  {Yang}}, \bibinfo {author} {\bibfnamefont {B.}~\bibnamefont {Gong}}, \ and\
  \bibinfo {author} {\bibfnamefont {W.}~\bibnamefont {Cui}},\ }\href {\doibase
  10.1103/PhysRevA.97.012119} {\bibfield  {journal} {\bibinfo  {journal} {Phys.
  Rev. A}\ }\textbf {\bibinfo {volume} {97}},\ \bibinfo {pages} {012119}
  (\bibinfo {year} {2018})}\BibitemShut {NoStop}%
\bibitem [{\citenamefont {Sontag}(2013)}]{sontag2013mathematical}%
  \BibitemOpen
  \bibfield  {author} {\bibinfo {author} {\bibfnamefont {E.~D.}\ \bibnamefont
  {Sontag}},\ }\href@noop {} {\emph {\bibinfo {title} {Mathematical control
  theory: deterministic finite dimensional systems}}},\ Vol.~\bibinfo {volume}
  {6}\ (\bibinfo  {publisher} {Springer Science \& Business Media},\ \bibinfo
  {year} {2013})\BibitemShut {NoStop}%
\bibitem [{\citenamefont {Raue}\ \emph {et~al.}(2009)\citenamefont {Raue},
  \citenamefont {Kreutz}, \citenamefont {Maiwald}, \citenamefont {Bachmann},
  \citenamefont {Schilling}, \citenamefont {Klingm{\"u}ller},\ and\
  \citenamefont {Timmer}}]{10.1093/bioinformatics/btp358}%
  \BibitemOpen
  \bibfield  {author} {\bibinfo {author} {\bibfnamefont {A.}~\bibnamefont
  {Raue}}, \bibinfo {author} {\bibfnamefont {C.}~\bibnamefont {Kreutz}},
  \bibinfo {author} {\bibfnamefont {T.}~\bibnamefont {Maiwald}}, \bibinfo
  {author} {\bibfnamefont {J.}~\bibnamefont {Bachmann}}, \bibinfo {author}
  {\bibfnamefont {M.}~\bibnamefont {Schilling}}, \bibinfo {author}
  {\bibfnamefont {U.}~\bibnamefont {Klingm{\"u}ller}}, \ and\ \bibinfo {author}
  {\bibfnamefont {J.}~\bibnamefont {Timmer}},\ }\href {\doibase
  10.1093/bioinformatics/btp358} {\bibfield  {journal} {\bibinfo  {journal}
  {Bioinformatics}\ }\textbf {\bibinfo {volume} {25}},\ \bibinfo {pages} {1923}
  (\bibinfo {year} {2009})},\ \Eprint
  {http://arxiv.org/abs/https://academic.oup.com/bioinformatics/article-pdf/25/15/1923/16889623/btp358.pdf}
  {https://academic.oup.com/bioinformatics/article-pdf/25/15/1923/16889623/btp358.pdf}
  \BibitemShut {NoStop}%
\bibitem [{\citenamefont {Najfeld}\ and\ \citenamefont
  {Havel}(1995)}]{najfeld1995derivatives}%
  \BibitemOpen
  \bibfield  {author} {\bibinfo {author} {\bibfnamefont {I.}~\bibnamefont
  {Najfeld}}\ and\ \bibinfo {author} {\bibfnamefont {T.~F.}\ \bibnamefont
  {Havel}},\ }\href {https://doi.org/10.1006/aama.1995.1017} {\bibfield
  {journal} {\bibinfo  {journal} {Advances in applied mathematics}\ }\textbf
  {\bibinfo {volume} {16}},\ \bibinfo {pages} {321} (\bibinfo {year}
  {1995})}\BibitemShut {NoStop}%
\bibitem [{\citenamefont {Li}\ \emph {et~al.}(2018)\citenamefont {Li},
  \citenamefont {Pezz{\`e}}, \citenamefont {Gessner}, \citenamefont {Ren},
  \citenamefont {Li},\ and\ \citenamefont {Smerzi}}]{e20090628}%
  \BibitemOpen
  \bibfield  {author} {\bibinfo {author} {\bibfnamefont {Y.}~\bibnamefont
  {Li}}, \bibinfo {author} {\bibfnamefont {L.}~\bibnamefont {Pezz{\`e}}},
  \bibinfo {author} {\bibfnamefont {M.}~\bibnamefont {Gessner}}, \bibinfo
  {author} {\bibfnamefont {Z.}~\bibnamefont {Ren}}, \bibinfo {author}
  {\bibfnamefont {W.}~\bibnamefont {Li}}, \ and\ \bibinfo {author}
  {\bibfnamefont {A.}~\bibnamefont {Smerzi}},\ }\href {\doibase
  10.3390/e20090628} {\bibfield  {journal} {\bibinfo  {journal} {Entropy}\
  }\textbf {\bibinfo {volume} {20}} (\bibinfo {year} {2018}),\
  10.3390/e20090628}\BibitemShut {NoStop}%
\end{thebibliography}%

\clearpage

\appendix
\section{Quadratic time dependence in the Fisher Information}\label{AppendixFI}
The quadratic growth in the Fisher Information (FI) with pulse duration can be derived straightforwardly in the case of a term, $g_i h_i$ with $g_i\in{\bf g}$, appearing in a time-independent Hamiltonian, $H=...+g_i h_i+...$.  Then $P_0(T)=|\psi_0(T)|^2$, where $\psi_0(T)=\bra{0}U(T)\ket{0}$ and $U(t)=e^{-itH}$.  Using an integral expression for derivatives of operator exponentials (see equation 13 of \citet{najfeld1995derivatives}), we find
\begin{align}
    \partial_{g_i}P_0(T) &= 2\, {\rm Re}\big[\psi_0^*(T)\bra{0}\partial_{g_i}U(T)\ket{0}],\nonumber\\
    &= 2\,{\rm Re}\big[\psi_0^*(T)\bra{0}U(T)\times\nonumber\\
    &\hspace{1.2cm}\int_0^1 d\tau\, U^\dagger(\tau T)\partial_{g_i}(-i T H) U(\tau T)\ket{0}\big]\nonumber\\
    &= 2 \,T\,{\rm Im}\big[\psi_0^*(T)  \bra{0}U(T) \bar h_i \ket{0}\big],\label{eqn:dP0dt}
\end{align}
where $\bar h_i = \int_0^1 d\tau\, h_i(\tau T) $ with $h_i(t)\equiv U^\dagger(t)  h_i U(t)$  is the time-averaged expectation of $h_i(t)$ on the interval \mbox{$t\in[0,T]$}.  In the last line,  the term in square-brackets is an oscillatory, bounded function of $T$, and the prefactor is linear in $T$.  It  follows that $(\partial_{g_i}P_0(T))^2$ scales as $T^2$, and so $F=\big(\partial_{g_i}P_0(T)\big){}^2/(P_0-P_0^2) = T^2 \vartheta({\bf g},T)$, as stated in \cref{eqn:FIscaling}.  This is consistent with specific simulations shown in \cref{fig:paramscaling}(d), and in \cite{PhysRevA.92.032124}.  

This property generalises straightforwardly for time-dependent piecewise-constant (PWC) controls whose segment durations  scale with overall control duration. Briefly, we transform to a suitable rotating frame  and make a rotating-wave approximation to remove explicit time-dependence.  In this frame, the evolution is given by a concatenation of unitary-evolution operators, each generated by a time-independent Hamiltonian (in the rotating frame) that depends linearly on $g_i$.  The proportionality in \Cref{eqn:dP0dt} will hold for each segment, and using the product rule for differentiation, we see that the growth of the FI will then also be quadratic in the total duration.

\section{Alternative Cost Functions} \label{AppendixAltCost}

The $Q$-weighted  averaging over $m$ and $m'$ implicit in \cref{eqn:ave} gives the statistically-correct expectation value for the APC, but it also means that low-probability measurement outcomes are weakly represented.  This could lead to some fragility in OBSID if these rare outcomes have a posterior distribution with high covariance, or other pathologies such as having very few samples represented in the prior population.  We are free to modify the cost function to penalise the risk of such situations occurring. 

 We comment here briefly  on alternative cost functions that are more robust, or enjoy other advantages. While we do not use them in the results reported in the main body of this paper, it is conceptually useful to abstract away from the specific choice of purely information-maximising cost function.

One way to make the cost function more robust is to use different aggregate measures over $m$.  For example, a conservative `variance-averse' cost function  would track the worst-case  posterior covariance for any possible experimental measurement.  
In this case, the \emph{maximum} posterior covariance (MaxPC) cost would be
\begin{equation}
        C_{\rm MaxPC}({\bf c}^{(j)})={\max}_m
        \Tr\big(A\cdot {\Sigma}_{j}(m)\big)
\end{equation}
Clearly, a continuum of variations are possible using different norms to measure the distribution of $ {\Sigma}_{j}(m)$ over $m$ or the prior population $G_{j-1}$.

Other cost functions are also available based on approximations for the expected information gain  \cite{NEURIPS2019_d55cbf21,pmlr-v108-foster20a}. In general, cost functions may be derived based on different uncertainty estimators \cite{e20090628}.  We have developed a further alternative, based on a modified Fisher information, $C_{\rm MFI}$, which relies on both the Cramer-Rao bound as an estimate of the sensitivity of the response, $P_0$, and a bespoke penalty function designed to disfavour oscillatory responses.  For completeness we formualte this below.


The APC requires averaging over hypothetical measurement outcomes which is computationally intensive \cite{NEURIPS2019_d55cbf21,pmlr-v108-foster20a,PRXQuantum.3.020350}. Conversely, the MFI requires no such averaging, but does use gradients directly in the cost to evaluate the Fisher information, and in a penalty term that discounts  oscillatory responses. These gradients are provided by automatic differentiation, so add a constant computational cost.  However, optimising MFI with gradient-based optimisers introduces an additional round of  automatic differentiation, so that the cost gradients include second-order derivatives of the response, which add further computationally overhead as well.  Both cost functions therefore have significant computational overheads.  In both cases the computation is parallelisable over the prior population ${\bf g}_s\in G_{j-1}$, and so the computation benefits from high-performance multiprocessor computing resources.

\subsubsection*{Modified Fisher Information Cost Function} 

Choosing control pulses that maximise the FI is a conceptually appealing approach to choosing maximally informative control pulses.  Given a prior distribution $\PDF_{j-1}({\bf g})$ over the system parameters, we seek to maximise the expected FI
\begin{equation}
    \bar {\rm FI}({{\bf c}})=\int d^p{\bf g} \,{\rm FI}({\bf g}; {{\bf c}})\,\PDF_{j-1}({\bf g}).
\end{equation}

However, if the evolution is unitary, and $g_i$ is a coefficient in the system Hamiltonian, then ${\rm FI}({\bf g}, T) \sim T^2 \vartheta({\bf g},T)$, where $\vartheta$ is some generally oscillatory function of pulse duration and the Hamiltonian parameters as shown in \cref{fig:paramscaling}.  Optimising $\bar F$ will then yield arbitrarily long pulse sequences.

This reflects the fact that the FI is a point-estimator of the sensitivity of a probability distribution to infinitesimal variations in a parameter, and this sensitivity can grow unboundedly (at the cost of longer pulse duration).

For parameter estimation however, we have a \emph{finite} prior uncertainty (illustrated as the white disk in \Cref{fig:paramscaling}), and so we have to discriminate amongst plausible model parameters distributed according to the prior $\Prior_{j-1}$.  If the pulse duration $T$ is too long, then the predicted response, $P_0({\bf g}, {\bf c})$, will typically be highly oscillatory over the support of $\PDF_{j-1}({\bf g})$, so a measurement of $\hat P_0( {\bf c})$  will not effectively constrain the posterior range of plausible parameter values ${\bf g}$, as illustrated in \Cref{fig:paramscaling}(c).  Instead, for a given prior covariance matrix $\Sigma_{j-1}$, the optimal pulse duration typically scales as $T^{(j)}\sim \Tr({\Sigma_{j-1}^{-1/2}})$, as discussed in \Cref{sec:FI}.

We therefore define a modified Fisher information (MFI) designed to penalise oscillatory behaviour in the response:
\begin{equation}
    C_{\rm MFI}({\bf c}, \Prior_{j-1})= -\bar {\rm FI}({{\bf c}})\big(1-\alpha \big\lfloor\Pfail({\bf c},\Prior_{j-1})-\beta\big\rfloor\big),
\end{equation}
where $\lfloor x\rfloor\equiv\max[x,0]$ is the ramp function, $\Pfail$ is a functional that penalises oscillatory responses, $\beta>0$ defines the tolerable penalty threshold, and $\alpha\sim 1$ scales the penalty. 

The prior distribution $\Prior_{j-1}$ is characterised by a prior mean $\bar {\bf g}_{j-1}$ and deviation matrix $\Sigma_{j-1}^{{1}/{2}}$ whose maximum  eigenvalue $ \lambda_{j-1}^{\maj}$ is associated to the  prior major uncertainty  direction $\hat {\boldsymbol{ \gamma}}_{j-1}^{\maj}$ in parameter space. 
A function that penalises oscillations in $P_0$ along the direction of $\hat {\boldsymbol{ \gamma}}_{j-1}^{\maj}$ is
\begin{equation}
    \Pfail({\bf c},\Prior_{j-1}) = \int d^p{\bf g} \,\PDF_{j-1}({\bf g})
     \Theta\big [-\tfrac
    {\hat {\boldsymbol{ \gamma}}_{j-1}^{\maj}\cdot\nabla_{\bf g} P_0({\bf g}; {{\bf c}})}
    {\hat {\boldsymbol{ \gamma}}_{j-1}^{\maj}\cdot\nabla_{\bf g} P_0(\bar{\bf g}_{j-1}; {\bf c})}\big],\nonumber
\end{equation}
where $\Theta[x]$ is the unit-step function. Qualitatively, if the gradient of the response, $\nabla_{\bf g} P_0$, projected along $\hat {\boldsymbol{ \gamma}}_{j-1}^{\maj}$ (i.e.\ the directional derivative of $P_0$), has a constant sign, either $+$ or $-$,   over the support of $\PDF_{j-1}({\bf g})$, then the penalty will be zero.  Changes in the sign of the directional derivative reveal oscillatory behaviour, and are penalised accordingly.

For a population sampled from the prior distribution ${\bf g}_s\sim \Prior_{j-1}$, we estimate the integrals in $\bar {F}$ and $\Pfail$ as discrete averages over the sample populations:
\begin{align}
    \bar {{\rm FI}}({{\bf c}})&=
    \tfrac{1}{S}{\sum}_{s=1}^S {\rm FI}({\bf g}_s; {{\bf c}}),\\
    \Pfail({\bf c},\Prior_{j-1}) &= 
    \tfrac{1}{S}{\sum}_{s=1}^S \Theta\big[-\tfrac
    {\hat {\boldsymbol{ \gamma}}_{j-1}^{\maj}\cdot\nabla P_0({\bf g}_s; {{\bf c}})}
    {\hat {\boldsymbol{ \gamma}}_{j-1}^{\maj}\cdot\nabla P_0(\bar{\bf g}_{j-1}; {\bf c})}\big].
\end{align}

We reiterate that these are implemented in a graph-based computational framework that allows for efficient automatic differentiation.  This allows us to compute the gradients that appear explicitly in $\bar {{\rm FI}}$ and $\Pfail$, and also to implement efficient  gradient-based optimisation of the cost function.


\section{Summary results of Alternative Experimental Pulse Parameterisations }\label{app:QCL_other_pulses}

\begin{figure}[t]
    {\includegraphics{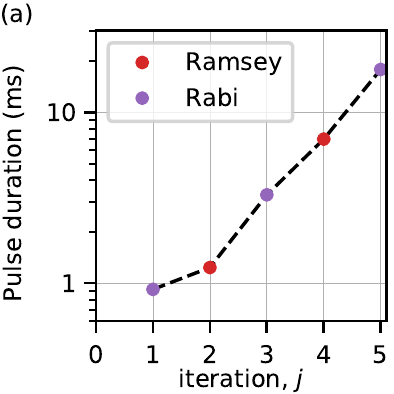}\hfill
    \includegraphics{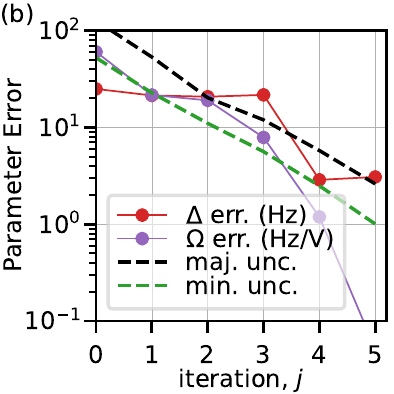}}
\caption{
Experimental OBSID results in a single-ion qubit calibration, using optimised Rabi or Ramsey pulse-types, as illustrated in \Cref{fig:RabiRamseyResults}a. 
(a) Pulse durations are shown at each iteration, with the autonomousluy-selected pulse-type indicated by colour. (b) The iterative improvement in the parameter accuracy, indicated by the major uncertainty of the posterior population, and verified by the absolute error relative to conventionally calibrated parameter values.  The floor in the absolute errors for $\Delta$ and $\Omega$ is set by the inferred uncertainty in the conventional calibration methods, which has a standard error of about $0.3$ Hz, for both $\Omega$ and $\Delta$.
}
\label{fig:QCLrabiramsey}
\end{figure}

As described in \Cref{sec:expQCLdemo} we have implemented other experimental control pulse sequences within the OBSID loop for calibration of the ion-trap experiment.  Specifically, these are (1) optimisable Rabi-Ramsey pulses, and (2) optimisable bang-bang control pulses.  Here we present the results of experimental OBSID with these pulse parameterisations.  

We note that the prior and posterior population plots look qualitatively similar to those shown in the main text, so we do not plot them here.

\subsubsection*{OBSID-selected Rabi-Ramsey Pulses}

\Cref{fig:QCLrabiramsey}(a)  shows the optimised pulse durations and types for the Rabi-Ramsey parameterisations illustrated in \cref{fig:RabiRamseyResults}(a).  The pulse type and the duration are autonomously chosen by OBSID for \emph{in-situ} experimental calibration of a single ion.  The pulse durations grow by a factor $\lesssim2$ per iteration.

\Cref{fig:QCLrabiramsey}(b) shows the progression in the absolute error (points) and the major uncertainty (dashed black line) with iteration count.  The major uncertainty provides a good estimate for the absolute error, although not a strict upper bound.  At the last iteration, $j=5$, the major uncertainty is \mbox{$\pm3$ Hz}, consistent with the absolute error in $\Delta$.  

The experimental results from the OBSID-selected Rabi-Ramsey sequence of pulses  are very similar to those in the  equivalent simulated results shown in \Cref{fig:RabiRamseyResults}.  In particular, at $j=5$ comparing \Cref{fig:RabiRamseyResults} and \Cref{fig:QCLrabiramsey}, we see that the pulse duration has grown by a factor of about 20 to 30 relative to the initial duration, and the major uncertainty has reduced by a similar factor.

\begin{figure}[t!]
\begin{minipage}[lt]{8.4cm}
    \begin{flushleft}
        \includegraphics[ trim=0 17 0 0, clip 
        ]{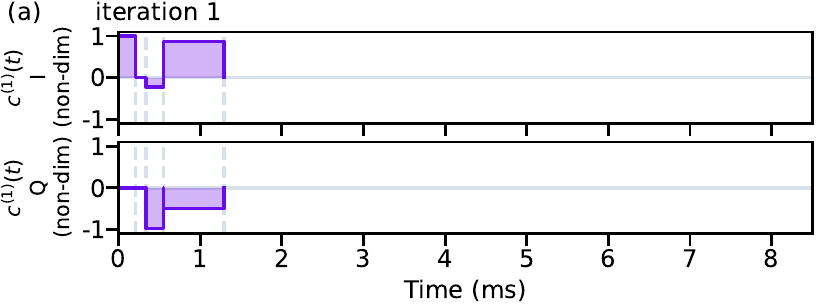}
        
        \vspace{1.7mm}
        \includegraphics
         [ trim=0 17 0 0, clip]{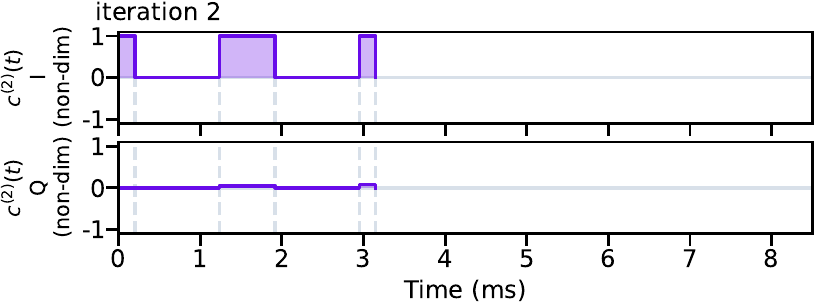}
         
         \vspace{1.7mm}
         \includegraphics[ trim=0 0 0 0, clip]{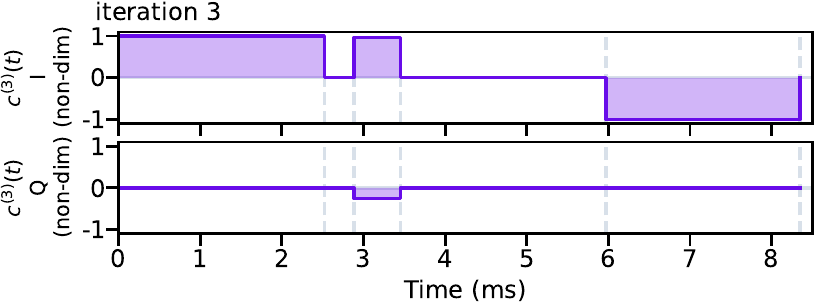}
         
        \vspace{1.7mm}
         {\hspace{1mm}
         \includegraphics{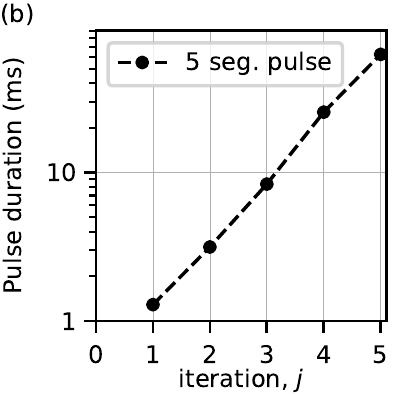}
         \includegraphics{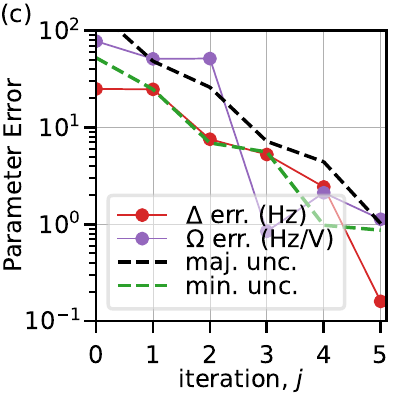}}
         \end{flushleft}
    \end{minipage}

\caption{
Experimental OBSID results in a single-ion qubit calibration, using optimised 5-segment bang-bang pulses, in which the control is either `on' (unit magnitude) with optimised phases,  or `off' (zero magnitude). (a) The sequence of complex-valued pulses (represented as in-phase [I] and quadrature [Q] components) generated in the first three iterations of OBSID, which optimises over segment durations and phases.  
(b) Pulse durations for each iteration, and (c)  the iterative improvement in the parameter accuracy.  The floor in the absolute errors for $\Delta$ and $\Omega$ is set by the inferred uncertainty in the conventional calibration methods, which has a standard error of about $0.3$ Hz, for both $\Omega$ and $\Delta$.
}
\label{fig:QCLbangbang}
\end{figure}

\subsubsection*{OBSID-selected Bang-Bang Pulses}

Bang-bang control involves repeatedly switching a control field  between an `on' state, with unit magnitude, and an  `off' state, with zero magnitude.  This is implemented here as a PWC with varying segment phases and  durations, and allowing the carrier phase to vary from segment to segment. 

This control type may be useful if there is a nonlinear transfer function between the control field generator (which in this experiment is an AWG) and the target system (which in this experiment is an ion).  In this case, fixing the `on'-state magnitude removes the need to fully calibrate the non-linear transfer function for a continuum of different pulse magnitudes.    

\Cref{fig:QCLbangbang}(a) shows the optimised complex-valued bang-bang pulses (represented as in-phase [I] and quadrature [Q] pairs), for the first three iterations of OBSID.  In this example, the control pulses are five-segment PWCs, and the first segment is always `on', with zero phase.  The segment durations are optimised independently with a lower bound of zero.

\Cref{fig:QCLbangbang}(b) shows that in this experimental run, the pulse durations grow by a factor of 2 to 3 per iteration, up to a maximum duration of about 60 ms at OBSID iteration $j=5$.  

The final pulse duration is somewhat longer in this example than the other experimental single-qubit  pulse parameterisations reported in Figures \ref{fig:QCLResults} and \ref{fig:QCLrabiramsey}.   There is a corresponding improvement in the final major uncertainty, which is shown in 
\Cref{fig:QCLbangbang}(c).  We see that at iteration $j=5$, the major uncertainty is \mbox{$\pm1$ Hz}, which is close to the \mbox{$\pm0.3$ Hz} floor set by standard experimental calibration methods used to independently determine the `true' parameter values for the system being measured.  The absolute error in the parameters remains consistent with the major uncertainty as an estimated bound.

\section{Practical System Identifiability}\label{sec:identifiability}

There is a more subtle issue, related to statistical identifiability \cite{sontag2013mathematical}, which we first describe, and then illustrate with further simulations.  The goal of OBSID is to accurately approximate the true parameter values ${\bf g}_\true$ that determine the system dynamics.  At each iteration, OBSID improves its estimate by optimally discriminating between different plausible values ${ \bf g}_s$ in a prior sample population, and it achieves this by choosing pulses that minimise the posterior uncertainty. 
However, for some control-pulse parameterisations, there may be a manifold of  model parameter values that give statistically close measurement results for \textit{all} available pulses within the  parameterisation.  In this case, the OBSID protocol will eventually stall, with a distribution whose major uncertainty is independent of the pulse duration.  At this point, OBSID will simply report a nearly stationary posterior parameter distribution. 

In control theory and statistics, a parameterised probability distribution is called \textit{identifiable} if empirical observations allow the parameters of the distribution to be identified uniquely, with no limitation on the number of observations allowed.  If more than one set of parameters yields the same distribution, then it is not \emph{structurally identifiable} \cite{10.1093/bioinformatics/btp358}.  In some cases, the response may be very shallow along some sub-manifold of parameter space, in which case the parameters may not be \emph{practically identifiable} \cite{10.1093/bioinformatics/btp358} given a finite number of measurements; this is the cause of OBSID stalling.

Practically, OBSID is limited to a finite number of empirical observations, and so practical statistical identifiability asks: for a given control pulse parameterisation, and  an upper limit on the number of measurements, $r$, 
can we improve the posterior parameter uncertainty by increasing the pulse duration (while tuning other control parameters, e.g.\ segment phases or amplitudes)?  As we now demonstrate, there are poorly parameterised control pulses for which the model is not practically identifiable.

\begin{figure}
\begin{flushleft}
{\includegraphics[
]{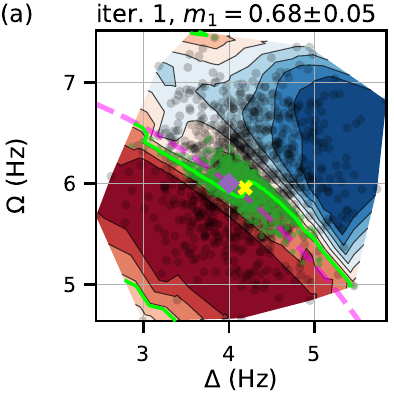}
\includegraphics[
]{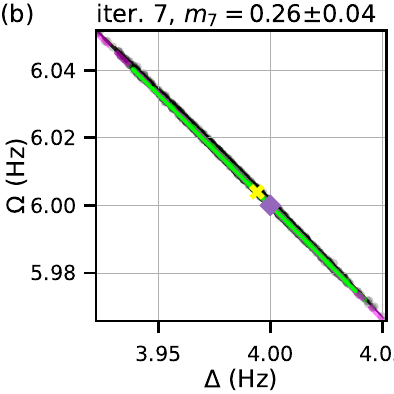}\hfill
\raisebox{8mm}{\includegraphics[]
{figs/labelled_legend.pdf}}

\includegraphics[width=4cm]
{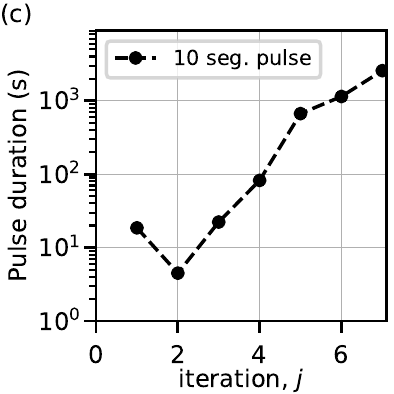}
\includegraphics[width=4cm]
{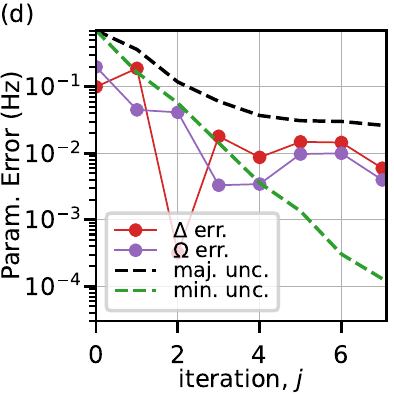}\hfill}

\caption{
A control example where OBSID fails to improve parameter estimates.  The control pulse has uniform amplitude, but with $n=10$ tunable phase segments $\phi_n(t)$, so that $c(t)=e^{i\phi_n(t)}$.  (a) and (b) show the prior populations and posterior populations  after the first  and the seventh iterations, respectively.  The magenta line indicates the contour $\Delta^2+\Omega^2=\Delta_\true^2+\Omega_\true^2$, showing that the control successfully collapses the posterior distribution onto the contour, but fails to constrain the posterior distribution along the contour tangent direction.  (c) shows that OBSID chooses increasingly long pulses, as in previous examples, however (d) shows that improvement in the major uncertainty `stalls' at a marginal uncertainty of around $\pm0.02$ Hz.
} 
\label{OBSIDfail}
\end{flushleft}
\end{figure}


An example of this effect is a parameterisation with a smaller number of free control parameters than the number of unknown system parameters, $p$.  For example, with a finite number of observation repetitions, $r$,  only using  Rabi pulses with only a controllable duration parameter, $T^{\rm (opt)}_{\rm Rabi}$, to identify the two parameters $\{\Delta,\Omega\}$ in \cref{eqn:H1q}, yields statistically significant information only about the combination $\Delta^2+\Omega^2$. 
Consider two `nearby' models in parameter space, described by ${\bf g}_a=\{\Delta_a,\Omega_a\}$ and ${\bf g}_b=\{\Delta_b,\Omega_b\}$, with $\Delta_a\approx\Delta_b$ and $\Omega_a\approx\Omega_b$.  These will give almost identical predictions for the Rabi response whenever $\Delta_a^2+\Omega_a^2=\Delta_b^2+\Omega_b^2$.  That is, within the 2-dimensional parameter space, there is 1-dimensional sub-manifold of models constrained by $\Delta^2+\Omega^2=\Delta_\true^2+\Omega_\true^2$ that are statistically indistinguishable, given a finite $r$.  It follows that within this sub-manifold, $\Delta$ and $\Omega$ cannot be independently determined.  The model is not practically identifiable with `Rabi-pulse-only' control.

We illustrate practical non-identifiability with a generalisation of the `Rabi-pulse-only' example.   Suppose that we attempt to use OBSID to identify the single-qubit, two-parameter model in \cref{eqn:H1q}, using a pulse with uniform amplitude, but with controllable, time-dependent phases, so that $c(t)=e^{i\phi_n(t)}$, where the control phase $\phi_n(t)$ is a real-valued PWC with $n\geq2$ segments.  For this  control parameterisation, there are more control parameters available than unknown model parameters, which is a necessary condition for OBSID to succeed.   Figures \ref{OBSIDfail}(a) and (b) show the prior and posterior population samples at the first and last of seven OBSID iterations; the final distribution is confined to a (nearly) 1-dimensional manifold aligned with the contour \mbox{$\Delta^2+\Omega^2=\Delta_\true^2+\Omega_\true^2$}, indicated by the dashed magenta line.  This posterior distribution is strongly covariant, so that neither parameter can be accurately fixed.  \Cref{OBSIDfail}(c) shows that the optimally chosen pulses still grow in duration, however \Cref{OBSIDfail}(d) illustrates that the major uncertainty saturates around iteration $j=4$. Further iterations decrease the major uncertainty of this distribution slowly as $1/\sqrt{r}$, rather than quickly as $1/T^{(j)}$.   This is a symptom that this `phase-only' control parameterisation is unable to practically identify the system \cite{10.1093/bioinformatics/btp358}.

This example shows how OBSID can fail for poorly-adapted control parameterisations that are unable to identify the system in practice.  Importantly, it also illustrates a symptom of this failure, which  is that major uncertainty stalls after several iterations of OBSID.  The resolution is  to  extend the control parameterisation to access more of the model parameter space.  This self-diagnostic capability is a key advantage of the protocol. 
In the example described above, including  control pulse amplitudes in the control parameterisation would suffice.

\end{document}